\documentclass[12pt]{article}

\usepackage{amssymb}
\usepackage{amsmath}
\usepackage[dvips]{graphicx}

\newcommand{\cF}{\mathcal{F}}

\newcommand{\we}{\wedge}

\newcommand{\om}{\omega}

\newcommand{\bea}{\begin{eqnarray}}
\newcommand{\eea}{\end{eqnarray}}
\newcommand{\be}{\begin{equation}}
\newcommand{\ee}{\end{equation}}
\newcommand{\ba}[1][rcl]{\begin{array}{#1}}
\newcommand{\ea}{\end{array}}

\newcommand{\nn}{\nonumber}

\newcommand{\half}{\frac12}
\newcommand{\inv}{^{-1}}
\newcommand{\wg}{\wedge}
\newcommand{\pr}{\partial}
\newcommand{\lpl}{\square_4}
\newcommand{\sg}{\sqrt{-G}}
\newcommand{\zt}{\tilde{z}}
\newcommand{\wt}{\tilde{w}}
\newcommand{\mt}{\tilde{m}}

\newcommand{\e}{\mathrm{e}}

\newcommand{\vol}{\mathrm{vol}}

\begin{document}

\title{On Normal Modes of a Warped Throat}
\author{\small Marcus K. Benna${}^{\,a}$,  Anatoly Dymarsky${}^{\,b}$, Igor R. Klebanov${}^{\,a,c}$
and Alexander Solovyov${}^{\,a,d}$}
\date{}
\maketitle
\thispagestyle{empty}
\begin{center}
\itshape
${}^{a}$ Department of Physics, Princeton University, Princeton, NJ  08544
\\[2.5mm]
${}^{b}$ Stanford Institute for Theoretical Physics,
Stanford, CA 94305
\\[2.5mm]
${}^{c}$ Princeton Center for Theoretical Physics, Princeton University, Princeton, NJ  08544
\\[2.5mm]
${}^{d}$ Bogolyubov Institute for Theoretical Physics,
Kiev 03680, Ukraine
\\[2.5mm]
\end{center}

\vspace{-10cm}
\begin{flushright}
PUPT-2253\\
{SU-ITP-07/25}\\
ITEP-TH-79/07
\end{flushright}
\vspace{10cm}

\abstract{ As shown in arXiv:hep-th/0405282, the warped deformed conifold has two bosonic
massless modes, a
pseudoscalar and a scalar, that are dual to the phase and the
modulus of the baryonic condensates in the cascading gauge theory.
We reconsider the scalar mode sector, mixing fluctuations of the
NS-NS 2-form and the metric, and include non-zero 4-d momentum
$k_\mu$. The resulting pair of coupled equations produce a
discrete spectrum of $m_4^2=- k_\mu^2$ which is interpreted as the
spectrum of $J^{PC}= 0^{+-}$ glueballs in the gauge theory. Similarly, we
derive the spectrum of certain pseudoscalar glueballs with $J^{PC}= 0^{--}$, which
originate from the decoupled fluctuations of the RR 2-form. We argue that each of the
massive scalar or pseudoscalar modes we find belongs to a 4-d massive axial vector or vector
supermultiplet.
We also discuss our results in the context
of a finite length throat embedded into a
type IIB flux compactification.
}

\newpage

\section{Introduction}
\label{sec:intro}

Duality between the cascading $SU(k(M+1))\times SU(kM)$ gauge theory and
type IIB strings on the warped deformed conifold \cite{KS} provides a rich yet
solvable example of gauge/string correspondence
\cite{Maldacena:1997re,Gubser:1998bc,Witten:1998qj}. For earlier work leading up to
this duality, see \cite{Klebanov:1998hh, Gubser:1998fp,KN, KT}, and for reviews
\cite{Herzog:2001xk,Strassler:2005qs}.
This background demonstrates in a geometrical language such
features of the $SU(M)$ supersymmetric gluodynamics as  color
confinement and the breaking of the $Z_{2M}$ chiral R-symmetry
down to $Z_2$ via gluino condensation \cite{KS}. In fact, it has been argued \cite{KS}
that by reducing the continuous parameter $g_s M$ one can
interpolate between the cascading theory solvable in the
supergravity limit and ${\cal N}=1$ supersymmetric $SU(M)$ gauge
theory.

The problem of finding the spectra of bound states at large $g_s
M$ can be mapped to finding normalizable fluctuations around the
supergravity background. This problem is complicated by the
presence of 3-form and 5-form fluxes, but some
results on the spectra are already available in the
literature \cite{Krasnitz,Caceres,GHK,BHM1,BHM2,Argurio,Dymarsky}. A
particularly impressive effort was made by Berg, Haack and M\" uck
(BHM) who used a generalized PT ansatz \cite{PT} to derive and
numerically solve a system of seven coupled scalar equations
\cite{BHM1,BHM2}. Each of the resulting glueballs is even under
the charge conjugation $Z_2$ symmetry preserved by the KS solution
(this symmetry was called the ${\cal I}$-symmetry in \cite{GHK}), and
therefore has $J^{PC}= 0^{++}$.
The present paper will study three other families of glueballs,
which are odd under the ${\cal I}$-symmetry.  Two of them
originate from a pair of coupled scalar equations, generalizing the zero
momentum case studied in \cite{GHK}, and have $J^{PC}= 0^{+-}$.
The third, pseudoscalar family arises from a
decoupled fluctuation of the RR two-form $C_2$ and has $J^{PC}= 0^{--}$.

An important aspect of the low-energy dynamics
is that the baryonic $U(1)_B$ symmetry is broken
spontaneously by the condensates of baryonic operators $\mathcal{A}$ and $\mathcal{B}$.
This phenomenon, anticipated in the cascading gauge theory in
\cite{KS,Aharony}, was later demonstrated on the supergravity
side where the fluctuations corresponding to the pseudoscalar
Goldstone boson and its scalar superpartner \cite{GHK}, as well as the fermionic
superpartner \cite{Argurio}, were identified.
Furthermore, finite deformations along the scalar direction give
rise to a continuous family of supergravity solutions
\cite{Butti,DKS,Benna} dual to the baryonic branch,
$\mathcal{A}\mathcal{B} = {\rm const}$, of the gauge theory moduli
space.

The main purpose of the present paper is to obtain a deeper
understanding of the GHK scalar fluctuations \cite{GHK} and their radial excitations.
Our motivation is two-fold. On the one hand, we seek an improved understanding of the glueball
spectra and their supermultiplet structure. On the other, we would like to shed new light on the normal
modes of the warped deformed conifold
throat embedded into a string compactification, which has played a role in
models of moduli stabilization \cite{KKLT} and D-brane inflation \cite{KKLMMT,BDKMS}.
In such inflation models, the
reheating of the universe involves emission of
modes localized near the bottom of the throat, which are dual to glueballs in the gauge
theory \cite{Barnaby,Kofman,Frey}.

This paper is structured as follows. In section 2 we construct a
generalization of the ansatz for the NSNS 2-form and metric
perturbations that allows us to study radial excitations of the
GHK scalar mode. We derive a system of coupled radial
equations and determine their spectrum (the details of the
numerical treatment are presented in Appendix~\ref{app:numerics}). In section 3 we
show that a similar ansatz for the RR 2-form perturbation
decouples from the metric giving rise to a single decoupled
equation for pseudoscalar glueballs. In section 4 we argue that the scalar glueballs we find
belong to massive axial vector multiplets, and the pseudoscalar glueballs belong to
massive vector multiplets. Agreement of the corresponding
equations is explicitly demonstrated
in the large radius (KT) limit. In section 5 we give a
perturbative treatment of the coupled equations for small mass
that allows us to study the scalar mass in models where
the length of the throat is finite. Review of the supergravity equations and of the
warped deformed conifold, as well as some technical details, are delegated to the Appendices.

\section{Radial Excitations of the GHK scalar}

\label{sec:radial_excitations}
The ansatz
that produced a normalizable scalar mode independent of the four-dimensional coordinates
$x^\mu$ was \cite{GHK}
\be \label{ghkansatz}
\delta B_2 = \chi(\tau)\, dg^5\ , \qquad
\delta G_{13} = \delta G_{24} = \psi (\tau)
\ .
\ee
Our first goal is to find a generalization of this ansatz that
will allow us to study the radial excitations of this massless scalar, i.e.~the series of modes that exist at
non-vanishing $k_\mu^2 =- m_4^2$.
Thus, we must include the dependence of all fields on $x^\mu$.
Such an ansatz that decouples from other fields at linear order is
\be\label{ansatz}
\ba
\delta F_3 &=& 0 \,,
\\
\delta F_5 &=& 0 \,,
\\
\delta B_2 &=& \chi(x,\tau)\, dg^5 + \pr_\mu \sigma(x,\tau)\, dx^\mu \wg g^5 \,,
\\
\delta H_3 \equiv d \delta B_2 &=& \chi^\prime\, d\tau \wg dg^5 +
\pr_\mu (\chi-\sigma)\, dx^\mu \wg dg^5 + \pr_\mu \sigma^\prime\, d\tau \wg dx^\mu \wg g^5 \,,
\\
\delta G_{13} = \delta G_{24} &=& \psi (x,\tau) \,. \ea\ee
The ansatz for $\delta B_2$ originates from the longitudinal component of a 5-d vector:
\be
\delta B_2= (A_\tau d\tau+ A_\mu dx^\mu)\wg g^5
\ .
\ee
Requiring the 4-d field strength to vanish, $F_{\mu\nu}=0$, restricts $A_\mu$
to be of the form $\partial_\mu$ acting on a function. Then, choosing
\be
A_\tau=-\chi'\ , \qquad A_\mu = \partial_\mu (\sigma - \chi)
\ ,
\ee
we recover the ansatz (\ref{ansatz}) up to a gauge transformation.

Yet another gauge equivalent way of writing (\ref{ansatz}) is
\be
\delta B_2 = (\chi-\sigma)\, dg^5\ -\sigma'\, d\tau \wg g^5\ .
\ee
The new feature of our ansatz compared to
the generalized PT ansatz used in \cite{BHM1,BHM2} is the presence of the second function
in $\delta B_2$ which multiplies $d\tau\wedge g^5$. Terms of this type, which are allowed by the
4-d Lorentz symmetry, turn out to be crucial for studying the modes that are odd under
the ${\cal I}$-symmetry.

Using $\delta(G\inv) = - G\inv\, \delta G\, G\inv$, we find that
$\delta G^{13} = \delta G^{24} = - G^{11}\, G^{33}\, \psi$.
The unperturbed metric components (see Appendix~\ref{app:KS_solution} for a review of the KS solution) are
\bea
G^{11} = G^{22} &=& \frac{2}{\epsilon^{4/3} K(\tau)\sinh^2 (\tau/2) h^{1/2}(\tau)} \,,
\\
G^{33} = G^{44} &=& \frac{2}{\epsilon^{4/3} K(\tau)\cosh^2 (\tau/2) h^{1/2}(\tau)} \,,
\\
G^{55} = G^{\tau\tau} &=& \frac{6\, K(\tau)^2}{\epsilon^{4/3} h^{1/2}} \,,
\\
G^{\mu\nu} &=& h^{1/2}\, \eta^{\mu\nu} \,.
\eea
In order to find the dynamic equations for the functions $\psi$,
$\chi$ and $\sigma$ in (\ref{ansatz}) we study the linearized
supergravity equations below (type IIB SUGRA equations are reviewed
in Appendix~\ref{app:SUGRA_eqns}).

\subsection{Equations of Motion for NSNS- and RR-Forms}

All the Bianchi identities are automatically satisfied with the
ansatz (\ref{ansatz}). Indeed, the relation $d\delta H_3 = 0$ is obvious, and consistent with vanishing  $dF_5$ we find that $\delta H_3 \wedge F_3=0$
(using eqs.~(\ref{F3}) and (\ref{dg5}) one can verify that $dg^5
\wg F_3 = 0$ and $d\tau \wg g^5 \wg F_3 =0$).

The self-duality equation for $F_5$ reads
\be
\delta \ast F_5 = 0 \,.
\ee
Given that $F_5$ has components along $g^1 \wg g^2 \wg g^3 \wg g^4
\wg g^5$ and along $d^4x \wg d\tau$, our adopted deformation of
the metric does not affect $\ast F_5$ to first order.

Even though the variations of the forms $F_3$ and $F_5$ are zero,
the deformations of their Hodge duals $\delta\ast F_3$ and
$\delta\ast F_5$ will in general be non-zero because of the
deformations of metric components. In the equation for $F_3$
\be
d\delta\ast F_3 = F_5 \wedge \delta H_3 \,,
\ee
the product $F_5 \wedge \delta H_3$ vanishes identically. From the explicit form of $F_3$ we see that
the Hodge dual of the  first two terms in (\ref{F3}) will be a closed form $\delta \ast F_3 = A(x,\tau)\, d^4x \wg
d\tau \wg (\ldots)$. The third term in
(\ref{F3}), $F' d\tau\wedge (g^1\wedge g^3 + g^2\wedge g^4)$, is
not affected by the deformation of the metric, and thus
$d\delta\ast F_3=0$ is satisfied identically.

The remaining equations are nontrivial. In particular
\be
d\delta\ast H_3 = 0 \,,
\ee
turns out to be more complicated than the equation for  $F_3$. The
variation
\be
\delta\ast H_3 = \ast \delta H_3 + \delta_G \ast H_3
\ee
consists of two parts: $\ast\delta H_3$ accounting for the deformation of the form $H_3$ itself, and $\delta_G \ast H_3$ arising from the deformation of the Hodge star. Explicit calculation shows that
\be\ba
\ast\delta H_3 &=& - \sg\, G^{11}\, G^{33}\, G^{55}\, \chi^\prime\, d^4x \wg dg^5 \wg g^5 - \sg\, G^{11}\, G^{33}\, |G^{\mu\mu}|\, \pr_\mu(\chi-\sigma)\, \ast_4 dx^\mu \wg d\tau \wg dg^5 \wg g^5
\\
&&
+ \half \sg\, (G^{55})^2\, |G^{\mu\mu}|\, \pr_\mu \sigma^\prime\, \ast_4 dx^\mu \wg dg^5 \wg dg^5 \,,
\\
\delta_G \ast H_3 &=& - \frac{g_s M \alpha^\prime}{2}\, \sg\,
G^{11}\, G^{33}\, G^{55} \bigl[ f^\prime G^{11}+k^\prime G^{33} \bigr] \psi\, d^4x \wg dg^5 \wg g^5 \,.
\ea\ee
The four-dimensional Hodge star $\ast_4$ is  taken w.r.t.\ the standard Minkowski metric.
Differentiating this expression for $\delta \ast H_3$ and equating to zero the coefficients multiplying linearly independent forms gives three equations:
\bea
\label{eq1} d^4x \wg dg^5 \wg dg^5 &:&\quad 2\, G^{11}\, G^{33}
\left[ \frac{g_s M \alpha^\prime}{2} \bigl[ f^\prime G^{11}+k^\prime G^{33} \bigr] \psi + \chi^\prime \right] = G^{55}\, h^\half\, \lpl \sigma^\prime \,,
\\
\nn d^4x \wg d\tau \wg dg^5 \wg g^5 &:&\quad \pr_\tau \left\{ \sg\, G^{11}\, G^{33}\,
G^{55} \left[ \frac{g_s M \alpha^\prime}{2} \bigl[ f^\prime G^{11}+k^\prime G^{33} \bigr] \psi + \chi^\prime \right] \right\} +
\\
\label{eq2} &&\quad + \sg\, G^{11}\, G^{33}\, h^\half\, \lpl (\chi-\sigma) = 0 \,,
\\
\label{eq3} \ast_4 dx^\mu \wg d\tau \wg dg^5 \wg dg^5 &:&\quad 2\, \sg\, G^{11}\, G^{33}\, h^\half\, \pr_\mu(\chi-\sigma) + \pr_\tau \left\{ \sg\, (G^{55})^2\, h^\half\, \pr_\mu \sigma^\prime \right\} =0 \, ,
\eea
where we have substituted for the warp factor $|G^{\mu\mu}|=h^\half$
(no summation over $\mu$ is implied).
Not all of these equations are independent. Indeed, using (\ref{eq1}) equation (\ref{eq2}) simplifies to
\be
\pr_\tau \left\{ \sg\, (G^{55})^2\, h^\half\, \lpl \sigma^\prime \right\} + 2\, \sg\, G^{11}\, G^{33}\, h^\half\, \lpl (\chi-\sigma) = 0 \,.
\ee
This is exactly what we obtain by acting on (\ref{eq3}) with $\pr^\mu$ and contracting indices. Thus only (\ref{eq1}) and (\ref{eq3}) are independent.
The coefficient functions in these equations are given by (we have dropped some inessential constant factor in $\sg$):
\bea
f^\prime G^{11}+k^\prime G^{33} &=&
\frac{2\, (\sinh 2\tau -2\tau)} {{\epsilon}^{4/3}\,\sqrt{h(\tau) }\,K(\tau)\, {\sinh^3\tau}} =
\frac{4\, K(\tau)^2} {{\epsilon}^{4/3}\,\sqrt{h(\tau) }\,} \,,
\\
G^{11}\, G^{33} &=& \frac{16}{{\epsilon}^{8/3}\,h(\tau) \,{K(\tau)}^{2}\,{\sinh^2\tau}} \,,
\\
\sqrt{h}\, G^{55} &=& \frac{6\,{K(\tau)}^{2}}{{\epsilon}^{4/3}}\,,
\\
\sg\, G^{11}\, G^{33}\, h^{1/2} &\sim& \frac{4}{K(\tau)^2} \,,
\\
\sg\, (G^{55})^2\, h^{1/2} &\sim& 9\, K(\tau)^{4}\,\sinh^2\tau \,.
\eea
Taking into account these expressions, equations (\ref{eq1}) and (\ref{eq3}) read
\bea
2 (g_s M\alpha^\prime)\,  \frac{K(\tau)^2} {{\epsilon}^{4/3}\,\sqrt{h(\tau) }\,}\, \psi +
\chi^\prime &=& \frac{3}{16}\, \epsilon^{4/3}\, h(\tau)\, K(\tau)^4\, \sinh^2\tau\, \lpl \sigma^\prime \,,
\\
\pr_\mu (\chi-\sigma) +\frac{9}{8}\, K(\tau)^2\, \pr_\tau \Bigl\{ K^4\, \sinh^2\tau\, \pr_\mu \sigma^\prime\Bigr\} &=& 0\,.
\eea

\subsection{Einstein Equations}
The first order perturbation of the Ricci curvature tensor is given by
\be
\delta R_{ij} = \frac{1}{2} \left( -{{\delta G_a}^a}_{;ij} -
{\delta G_{ij;a}}^a + {\delta G_{ai;j}}^a + {\delta G_{aj;i}}^a
\right) \ ,
\ee
where covariant derivatives and contractions of indices are performed using the unperturbed metric. The first term in
this expression vanishes because the metric perturbation is
traceless. The remaining three terms combine to give the
only non-zero perturbations $\delta R_{13} = \delta R_{24}$:
\begin{eqnarray} \label{Ricci}
\delta R_{13} &=&
-\frac{3}{\epsilon^{4/3}} K^3 \sinh(\tau) z
\left[ {K'' \over K} + {1 \over 2} {h'' \over h} + {z'' \over
z}
+ {(K')^2 \over K^2} - {1 \over 2} {(h')^2 \over h^2}
+ {K' \over K} {h' \over h} +  \right.
\nonumber \\
&&
\left. 2 {K' \over K} {z' \over z}+\coth \tau \left( {h' \over h} + 4 {K' \over K}
+ 2 {z' \over z} \right) + 2 - {1 \over \sinh(\tau)^2}
- {4 \over 9} {1 \over \sinh(\tau)^2 K^6} \right] -\half h(\tau) K \sinh(\tau)\, \lpl z
\nonumber \\
&=& -\frac{3}{\epsilon^{4/3}} K^3 \sinh\tau z
\left[
\frac{1}{2} \frac{\left( \left(K \sinh(\tau) \right)^2 (\ln h)' \right)' }{(K \sinh(\tau))^2}
+\frac{\left( \left( K \sinh(\tau) \right)^2 z' \right)' }
{(K \sinh(\tau))^2 z} \right.
\nonumber \\
&& \left. -\frac{2}{\sinh(\tau)^2} - \frac{8}{9} \frac{1}{K^6 \sinh(\tau)^2}
+ \frac{4}{3} \frac{\cosh(\tau)}{K^3 \sinh(\tau)^2}
\vphantom{\frac{\left( \left( K \sinh(\tau) \right)^2 z' \right)' }
{(K \sinh(\tau))^2 z}} \right]
-\half h(\tau) K \sinh\tau\, \lpl z\,,
\label{deltaR13}
\end{eqnarray}
where $z(x,\tau)$ is defined by
\begin{equation}
\label{ztaudef}
\psi (x,\tau)= h^{1/2} K\sinh (\tau)\, z (x,\tau)=
2^{-1/3} [\sinh(2\tau) - 2\tau]^{1/3} h^{1/2} z (x,\tau) \,.
\end{equation}

The source terms on the right hand side of the Einstein equation
$R_{ij}=T_{ij}$ (\ref{lEE}) are due to the deformations of the metric and $B_2$
form. It turns out that the only nontrivial deformations are those
with indices $13$ or $24$, with $\delta T_{13}= \delta T_{24}$. Say, for the $13$
component $\delta T_{13}$ we have the following contributions:
\bea
\nn \frac{1}{4}\, \delta_B ( H_{1ab}\, H_3{}^{ab} ) &=&
\frac{1}{4}\, [ H_{1ab}\, \delta H_3{}^{ab} + \delta H_{1ab}\,
H_3{}^{ab} \bigr]
\\
\nn &=& \frac{1}{2}\,  \bigl[ G^{11}\, H_{12\tau}\, \delta
H_{32\tau} + G^{33}\, \delta H_{14\tau}\, H_{34\tau} \bigr]\,
G^{55}
\\
&=& - \frac{1}{4}\,  (g_s M \alpha^\prime)\, G^{55}\,
\bigl[G^{11}\, f^\prime+G^{33}\, k^\prime\bigr]\, \chi^\prime \ ,
\\
{g_s^2\over 96}\, \delta_G(F_{1abcd}F_3{}^{abcd}) &=& \frac{g_s^2}{4}\, (G^{11})^2\, (G^{33})^2\, G^{55}\, (F_{12345})^2\, \psi \,,
\\
\nn \frac{1}{4}\, \delta_G (H_{1ab} H_3{}^{ab}) &=& \frac{1}{2}\, \bigl[ H_{135}\, H_{315}\, \delta G^{13}\, G^{55} + H_{12\tau}\, H_{34\tau}\, \delta G^{24}\, G^{\tau\tau} \bigr]
\\
\nn &=& \frac{1}{2} \bigl[ (H_{135})^2-H_{12\tau}\, H_{34\tau} \bigr]\, G^{11}\, G^{33}\, G^{55}\, \psi
\\
&=& \frac{1}{8}\, (g_s M\alpha^\prime)^2\, \left[{1 \over 4}(k-f)^2-f^\prime k^\prime\right]\,
G^{11}\, G^{33}\, G^{55}\, \psi \,,
\eea
\bea
\nn \frac{g_s^2}{4}\, \delta_G(F_{1ab} F_3{}^{ab}) &=& \frac{g_s^2}{2}\, \bigl[ F_{125}\, F_{345}\, \delta G^{24}\, G^{55} + F_{13\tau}\, F_{31\tau}\, \delta G^{31}\, G^{\tau\tau} \bigr]
\\
\nn &=& \frac{g_s^2}{2} \bigl[ (F_{13\tau})^2-F_{125}\, F_{345} \bigr]\, G^{11}\, G^{33}\, G^{55}\, \psi
\\
&=& \frac{1}{8}\, (g_s M\alpha^\prime)^2\, \bigl[ F^{\prime2}-F(1-F) \bigr]\,
G^{11}\, G^{33}\, G^{55}\, \psi\,,
\\
\nn -\frac{1}{48}\, \delta_G \bigl[G_{13} (H_{abc} H^{abc}+g_s^2 F_{abc} F^{abc})\bigr] &=& -\frac{1}{8}\,
(H^2+g_s^2 F^2)\, \psi
\\
\nn &=&-\frac{1}{32}\, (g_s M\alpha^\prime)^2\, G^{55}\, \bigl[ (G^{11})^2\, f^{\prime 2} +(G^{33})^2\, k^{\prime 2}
\\
\nn && + \half \, G^{11}\, G^{33}\, (k-f)^2 + (G^{11})^2\, F^2 +(G^{33})^2\, (1-F)^2
\\
&& + 2\,G^{11}\, G^{33}\, F^{\prime 2}\bigr] \psi\, .
\eea
Denoting
\be
\delta T_{13} = \Bigl[A_1(\tau)+A_2(\tau)\Bigr]\, \psi(x,\tau) + B(\tau)\, \chi^\prime(x,\tau) \,,
\ee
where $A_1$ stands for the contribution from $F_5$, we get
\bea
A_1(\tau) &=& \frac{3 (g_s M\alpha^\prime)^4}{2^{1/3} \epsilon^{20/3} h^{5/2}}\,
\frac{ (\tau \coth \tau -1)^2 [\sinh (2\tau) - 2\tau]^{4/3}}{\sinh^6 (\tau)} \,,
\\
A_2(\tau) &=& -\, \frac{3\, (g_s M\alpha^\prime)^2}{8\epsilon^4h^{3/2}\sinh^6\tau}\, \Bigl[ 3\,\cosh 4\tau  -8\tau\,\sinh 2\tau -8\tau^2\,\cosh 2\tau
-8\,\cosh 2\tau +16\tau^2+5 \Bigr]\,,
\\
B(\tau) &=& -3\, (g_s M\alpha^\prime)\, \frac{(\sinh 2\tau-2\tau)\, K(\tau)}{\epsilon^{8/3}\, h(\tau)\, \sinh^3\tau} \,.
\eea
Then eliminating $\chi^\prime$ with the help of (\ref{eq1}) yields
\bea\label{Tij}
\nn \delta T_{13} &=& \frac{3}{2^{2/3}}\, \frac{(g_s M\alpha^\prime)^4}{\epsilon^{20/3} h^2 }\,
\frac{ (\tau \coth \tau -1)^2 [\sinh 2\tau - 2\tau]^{5/3}}{\sinh^6\tau}\, z(\tau) +\frac{3}{8 \cdot 2^{1/3} } \, \frac{(g_s M \alpha^\prime)^2}{\epsilon^{4} h}\, \frac{(\sinh(2\tau)  -2\tau)^{1/3}}{ \sinh^{6}(\tau)} \times
\\
\nn &&
\times \Bigl[\cosh(4\tau)+ 8(1+\tau^2) \cosh(2\tau)-24\tau \sinh(2\tau)+ 16\tau^2-9 \Bigr]\, z(x,\tau)
\\
&& - \frac{9}{16}\, \frac{g_s M\alpha^\prime}{\epsilon^{4/3}}\, \frac{\sinh 2\tau-2\tau}{\sinh\tau}\, K^5\, \lpl \sigma^\prime(x,\tau)
\\
\nn &=& -\frac{3}{\epsilon^{4/3}} K^3 \sinh \tau \left[ -\half\,\frac{(h')^2}{h^2} + \frac{1}{2} \frac{h''}{h} + \frac{K'}{K} \frac{h'}{h} + \coth \tau \frac{h'}{h}
\right]\, z - \frac{9}{16}\, \frac{g_s M\alpha^\prime}{\epsilon^{4/3}}\, \frac{\sinh 2\tau-2\tau}{\sinh\tau}\, K^5\, \lpl \sigma^\prime \,.
\eea
As mentioned above the perturbations $\delta T_{13} = \delta
T_{24}$ are the only non-zero components of $\delta T_{ij}$.
Equating (\ref{Ricci}) and (\ref{Tij}) we obtain the final form of the linearized Einstein equation.

\subsection{Two Coupled Scalars}
Combining the equations for the field strengths and the Einstein equations we
have the system
\bea
\label{chi_el} (g_s M\alpha^\prime)\,  \frac{\sinh 2\tau -2\tau}
{{\epsilon}^{4/3}\,\sinh^2\tau}\, z + \chi^\prime &=&
\frac{3}{16}\, \epsilon^{4/3}\, h(\tau)\, K(\tau)^4\,
\sinh^2\tau\, \lpl \sigma^\prime \,,
\\ \label{chi_sigma}
\pr_\mu (\chi-\sigma) &=& -\frac{9}{8}\, K(\tau)^2\, \pr_\tau \Bigl\{ K^4\, \sinh^2\tau\, \pr_\mu \sigma^\prime\Bigr\} \,,
\\
\nn \frac{\left( \left( K \sinh\tau \right)^2 z' \right)' }
{(K \sinh\tau)^2 } + \frac{\epsilon^{4/3} h}{6\, K^2}\, \lpl z &=& \left(
\frac{2}{\sinh^2\tau} +\frac{8}{9} \frac{1}{K^6 \sinh^2\tau}
- \frac{4}{3} \frac{\cosh\tau}{K^3 \sinh^2\tau}
\right) z
\\
&& + \frac{3}{16}\, (g_s M\alpha^\prime)\, \frac{\sinh 2\tau-2\tau}{\sinh^2\tau}\, K^2\, \lpl \sigma^\prime \,.
\eea
Note that $\chi$ can be eliminated between
(\ref{chi_el}) and (\ref{chi_sigma}). Further, a change of variables
\bea
\tilde{z} &=& z K\sinh(\tau)\ ,\\
\tilde{w} &=& { \epsilon^{4/3} \over g_s M \alpha'} K^5\sinh(\tau)^2\sigma'\ ,
\eea
leads to a more symmetric pair of equations
\bea \label{CoupledScalarsz}
\tilde{z}''-{2\over \sinh^2 \tau}\tilde{z}+{\epsilon^{4/3}h \over
6K^2}\lpl\tilde{z} &=& {3 (g_sM\alpha')^2 \over 16 \epsilon^{4/3} }{\sinh 2\tau-2\tau\over
K^2\sinh^3\tau}\lpl \tilde{w}\ ,
\\ \label{CoupledScalarsw}
\tilde{w}''-{\cosh^2\tau+1\over \sinh^2 \tau}
\tilde{w}+{\epsilon^{4/3}h \over 6K^2}\lpl\tilde{w} &=& {8\over
9}{\sinh 2\tau-2\tau\over K^2 \sinh^3 \tau }\tilde{z}\,.
\eea
Introducing the dimensionless mass-squared $\mt^2$ according to
\be
\label{mass}
\tilde m^2 = m_4^2\, {2^{2/3} (g_s M \alpha')^2 \over 6\, \epsilon^{4/3}} \,,
\ee
we can rewrite the equations for $\zt$ and $\wt$ as
\bea
\label{eq1f}
\tilde{z}''-{2\over \sinh^2 \tau}\tilde{z}+
\tilde m^2 { I(\tau)\over K^2(\tau)}
\tilde{z} &=& \tilde m^2 {9 \over 4\cdot 2^{2/3}}\, K(\tau)\, \tilde{w}\ ,
\\
\label{eq2f}
\tilde{w}''-{\cosh^2\tau+1\over \sinh^2 \tau}
\tilde{w}+ \tilde m^2 { I(\tau)\over K^2(\tau)} \tilde{w} &=&
{16\over 9}\, K(\tau)\, \tilde{z} \,.
\eea
This is a system of coupled equations which defines the mass
spectrum of certain scalar glueballs with positive 4-d parity.
The natural charge conjugation symmetry of the KS background is the ${\cal I}$-symmetry, under
which these modes are odd.
Therefore, we assign $J^{PC}=0^{+-}$ to this family of glueballs.\footnote{
For comparison, the glueballs found in \cite{BHM1,BHM2} are $0^{++}$.
The glueballs whose spectrum comes from the minimal scalar equation \cite{Krasnitz} resulting from the
analysis of graviton fluctuations are $2^{++}$. The axial vector
$U(1)_R$ fluctuations
\cite{Dymarsky} give rise to $1^{++}$ glueballs whose masses are also determined by the
minimal scalar equation.}

In the massless case these equations lead to the GHK solution \cite{GHK}.
If we assume $\lpl =-k_\mu^2=m_4^2=0$, then there are two
solutions~\cite{GHK}, $\zt_1=\coth\tau$ and $\zt_2=\tau\coth\tau-1$. The
solution for $\tilde z$ which is non-singular at the origin is
$\tilde z= \tau\coth \tau -1$. Substituting it into the second
equation, we find
\be \label{zerolapl}
\tilde{w}''-{\cosh^2\tau+1\over \sinh^2 \tau}
\tilde{w}={16\over 9}\, K(\tau)\, (\tau\coth \tau -1)
\equiv -\, \frac{2^{2/3}\, 8}{9}\, I^\prime(\tau) \sinh\tau \,.
\ee
The two solutions of the homogeneous equation are
$\wt_1=1/\sinh\tau$ and $\wt_2=(\sinh2\tau-2\tau)/\sinh\tau$; both
of them are singular either at zero or at infinity. This means
that the regular solution of the inhomogeneous equation
is uniquely fixed. With the Wronskian
$W(\wt_1,\wt_2)=\wt_1\wt_2^\prime-\wt_1^\prime\wt_2=4$, we can
find a general solution
\be
\wt(\tau) = -\, \frac{2^{2/3}\, 8}{9} \Bigl\{ \wt_1(\tau) \bigl[ C_1-\int^\tau\!dx\, \frac{\wt_2(x)}{W(x)}\, I^\prime(x)\sinh x\bigr] + \wt_2(\tau) \bigl[ C_2 + \int^\tau\!dx\, \frac{\wt_1(x)}{W(x)}\, I^\prime(x)\sinh x\bigr] \Bigr \} \,.
\ee
Integrating by parts and choosing the particular homogeneous solution to make $\wt$ well behaved
at both zero and infinity we get
\be \label{ghkres}
\wt(\tau) = -\, \frac{2^{2/3}\, 8}{9}\, \frac{1}{\sinh\tau} \int_0^\tau\! dx\, I(x) \sinh^2 x \,.
\ee
Alternatively, if we start with equation (3.18) in the GHK paper~\cite{GHK}, which determines $f_2$, and
introduce $\tilde w = f_2(\tau) K^2 \sinh \tau$, we get (\ref{zerolapl}),
up to a rescaling of the right hand side. Thus, we can use the solution (3.25) of \cite{GHK}
for $f_2(\tau)$ to read off the result (\ref{ghkres}).

Let us also note that the non-zero $\wt$ in the zero momentum case
$k_\mu=0$ is not in contradiction with the GHK solution. This is because $\wt$ enters (\ref{ansatz}) only
through $\partial_\mu \sigma$ which is zero as long as the momentum vanishes.

\subsection{Numerical Analysis}

To determine the spectrum of glueballs in the field theory, we
need to solve the eigenvalue problem for $\tilde m^2$ in the
infinite throat limit. This system of equations (\ref{eq1f}), (\ref{eq2f}) does not seem amenable to analytical
solution and we employ a numerical approach to find the spectrum
of normalizable solutions. It is convenient to use
the determinant method, which generalizes the standard shooting
technique to a system of several coupled equations (see
\cite{BHM2}). The detailed description of the numerical analysis
as well as of the subtleties specific to the system
(\ref{eq1f}), (\ref{eq2f}) is given in
Appendix~\ref{app:numerics}. The result is that the spectrum
consists of two distinct series,  each with a quadratic growth of
$\mt_n^2$ for large $n$. These series are interpreted as the radial excitation spectra
of two different particles. The
lowest eigenvalues ($\mt^2<100$) for these spectra are shown in
Table~\ref{tab:eigenvals}.
\begin{table}[tb]
\begin{center}
\caption{\small Non-zero eigenvalues with $\mt^2<100$. There
are the two distinct spectra. Both spectra can be fitted by
quadratic polynomials in the eigenvalue number $n$ (the red line
in the plots).} \label{tab:eigenvals}
\begin{tabular}{cc}
Spectrum I & Spectrum II \\
\begin{tabular}[b]{||c|c||c|c||c|c||c|c||}
\hline
$n$ & $\mt_n^2$ & $n$ & $\mt_n^2$ & $n$ & $\mt_n^2$ & $n$ & $\mt_n^2$ \\
\hline
$1$ & 4.53 & $5$ & 19.1 & $9$ & 43.3 & $13$ & 76.9 \\
\hline
$2$ & 7.30 & $6$ & 24.4 & $10$ & 50.8 & $14$ & 86.7 \\
\hline
$3$ & 10.7 & $7$ & 30.1 & $11$ & 58.9 & $15$ & 97.1 \\
\hline
$4$ & 14.6 & $8$ & 36.4 & $12$ & 67.6 &  & \\
\hline
\end{tabular} &
\begin{tabular}[b]{||c|c||c|c||c|c||c|c||}
\hline
$n$ & $\mt_n^2$ & $n$ & $\mt_n^2$ & $n$ & $\mt_n^2$ & $n$ & $\mt_n^2$ \\
\hline
$1$ & 0.129 & $6$ & 8.06 & $11$ & 30.1 & $16$ & 65.1 \\
\hline
$2$ & 0.703 & $7$ & 11.2 & $12$ & 35.5 & $17$ & 73.9 \\
\hline
$3$ & 1.76 & $8$ & 15.0 & $13$ & 42.1 & $18$ & 83.3 \\
\hline
$4$ & 3.33 & $9$ & 19.3 & $14$ & 49.2 & $19$  & 93.3 \\
\hline
$5$ & 5.43 & $10$ & 24.1 & $15$ & 56.9 &  & \\
\hline
\end{tabular}
\\
\includegraphics[width=0.45\textwidth]{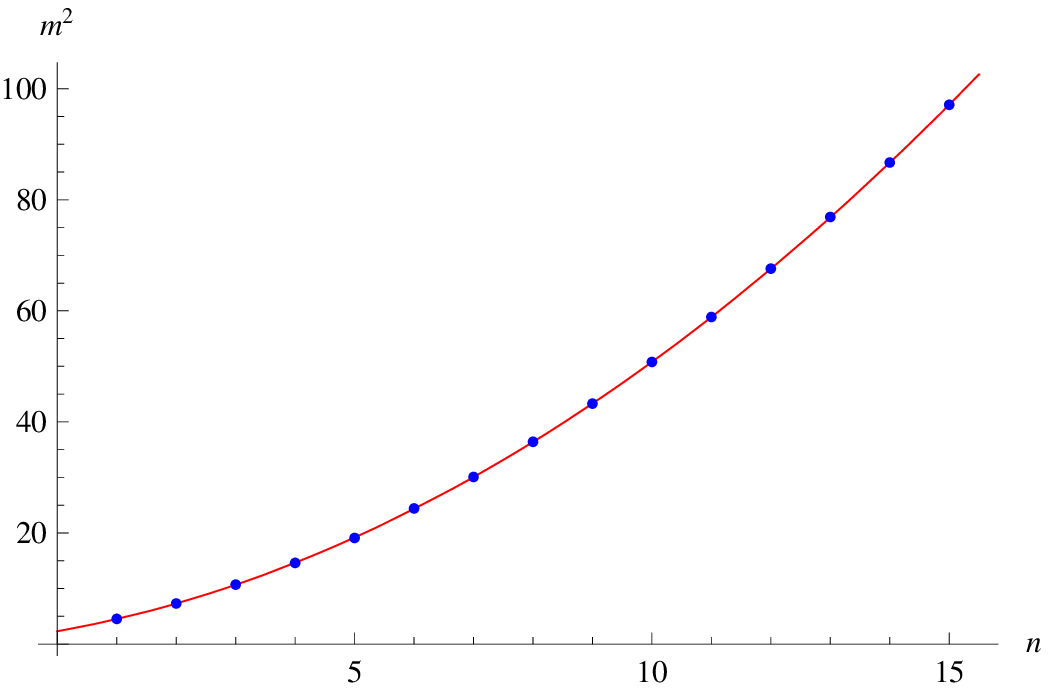} &
\includegraphics[width=0.45\textwidth]{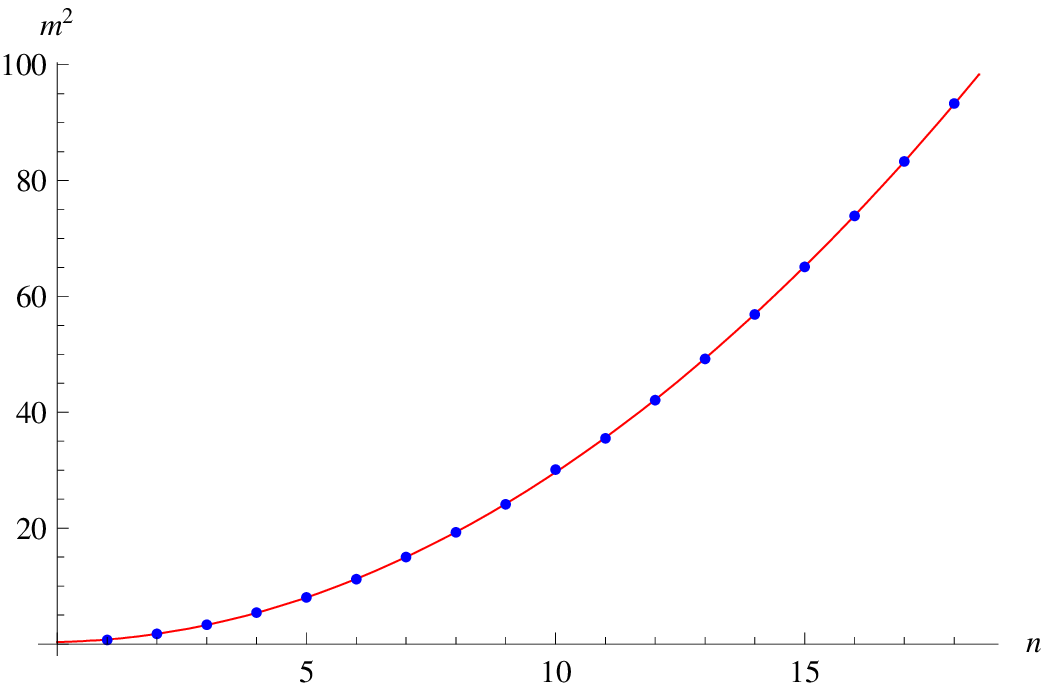}
\end{tabular}
\end{center}
\end{table}
The quadratic fit for spectrum I is
\bea
\tilde{m}_{In}^2 &=& 2.31 + 1.91\, n + 0.294\, n^2 \,.
\eea
For spectrum II (we drop the lowest eigenvalue when doing the fit)
\bea
\tilde{m}_{IIn}^2 &=& 0.36 + 0.14\, n + 0.279\, n^2 \,.
\eea
It is interesting to compare these results
with those found for the $0^{++}$
modes by Berg, Haack and M\" uck (BHM)~\cite{BHM2}.
The conventions of~\cite{BHM2} correspond to a particular choice of the KS parameters
(see Appendix~\ref{app:BHM}), and the relation between the masses is
\be
m_{BHM}^2 = (3/2)^{2/3} I(0)\, \mt^2 \approx 0.9409\,  \mt^2 \,.
\ee
Using this relation one can convert the mass eigenvalues to the BHM normalization.
We note that the lightest glueball we find, the first entry from spectrum II
in Table 1,
has $m_{BHM}^2\approx 0.121$. For comparison,
the lightest $0^{++}$ eigenvalue found in
\cite{BHM2} has $m_{BHM}^2\approx 0.185$. The fact that the $0^{+-}$ sector has the
lightest glueballs may be qualitatively understood as follows. Roughly speaking,
glueball masses increase with the dimensions of the operators that create them.
The lowest dimension operator from the $0^{++}$ sector is the gluino bilinear
${\rm Tr} \lambda \lambda$ of dimension 3, but the $0^{+-}$ sector contains an operator
of dimension 2, namely ${\rm Tr} ({\bar A} A - {\bar B} B)$.

Converting the asymptotics of the two spectra to BHM units, we find
\bea
m_{I\ BHM}^2 &\approx& 2.17 + 1.79\, n + 0.277\, n^2 \,,
\\
m_{II\ BHM}^2 &\approx& 0.34 + 0.13\, n + 0.262\, n^2 \,.
\eea
The coefficients of the quadratic terms are close to those found in \cite{BHM2}.
The quadratic dependence on $n$, which is characteristic of Kaluza-Klein theory, is a special
feature
of strongly coupled gauge theories that have weakly curved gravity duals (see \cite{Karch}
for a discussion).
Note that $m_4^2$ is obtained from ${\tilde m}^2$ through multiplying by a factor
$\sim T_s/(g_s M)$, where $T_s$ is the confining string tension.
Thus, for $n\ll \sqrt{g_s M}$ these modes are much lighter than the string tension scale, and
therefore much lighter than all glueballs with spin $>2$.
Such anomalously light bound states appear to be special to gauge theories that stay
very strongly coupled in the UV, such as the cascading gauge theory; they do not appear in
asymptotically free gauge theories. Therefore, the anomalously light
glueballs could perhaps be used as a `special signature' of gauge
theories with gravity duals if they are realized in nature.

One may be puzzled why the spectrum in Table~\ref{tab:eigenvals} does not include
the GHK massless mode. This is because in solving the coupled
equations (\ref{eq1f}), (\ref{eq2f}) we required that both
wave-functions $\tilde z$ and $\tilde w$
vanish as $\tau \to \infty$. This excludes the GHK zero mode which grows as
$\tilde z\sim \tau$. On the other hand, this growth is a lot slower than the exponential
growth found for generic solutions. The
meaning of the GHK mode as the baryonic branch modulus seems to be
well established since even the solutions at finite distance along
this modulus are available \cite{Butti,DKS}.
Thus, the GHK scalar zero-mode should be normalizable with a proper definition
of norm. In fact, the GHK pseudoscalar and its fermionic superpartner are
normalizable \cite{GHK,Argurio}; therefore, the supersymmetry of the problem implies
that the GHK scalar is normalizable as well and is part of the spectrum.

\section{Pseudoscalar Modes from the RR Sector}

\label{sec:another_scalar}
The type of ansatz used in section~\ref{sec:radial_excitations} works even more simply
for the RR 2-form field:
\be\label{ansatz2}
\ba
\delta H_3 &=& 0 \,,
\\
\delta F_5 &=& 0 \,,
\\
\delta C_2 &=& \chi(x,\tau)\, dg^5 + \pr_\mu \sigma(x,\tau)\, dx^\mu \wg g^5 \,,
\\
\delta F_3 \equiv d \delta C_2 &=& \chi^\prime\, d\tau \wg dg^5 + \pr_\mu (\chi-\sigma)\, dx^\mu \wg dg^5 + \pr_\mu \sigma^\prime\, d\tau \wg dx^\mu \wg g^5 \,.
\ea\ee
This ansatz is similar to, but somewhat simpler than the GHK pseudoscalar ansatz \cite{GHK} which involved
mixing with $\delta F_5$.
Since $\delta F_3\wedge H_3=0$, now it is consistent to set $\delta F_5=0$.
We also have $F_5\wedge \delta F_3=0$, so it is consistent to take
$\delta H_3=0$.
Finally, one needs to study mixing with metric fluctuations.
At a first glance it seems that $\delta G_{12}$ and $\delta G_{34}$ might need to be turned on,
but a more detailed analysis shows that their sources vanish:
\bea
\delta T_{12} &=& F_{13\tau} \delta F_2{}^{3\tau} + \delta F_{14\tau} F_2{}^{4\tau} \;=\; \frac{M\alpha^\prime}{2}\, G^{33}\, G^{55} \bigl[ F^\prime\chi^\prime -F^\prime\chi^\prime \bigr] \;=\; 0 \,,
\\
\delta T_{34} &=& F_{31\tau} \delta F_4{}^{1\tau} + \delta F_{32\tau} F_4{}^{2\tau} \;=\; 0 \,.
\eea
Thus, the perturbation (\ref{ansatz2}) decouples from all other modes, and
the only non-trivial linearized equation is
\be
d *\delta F_3=0 \,.
\ee
The calculation we need to perform is the same as in section~2.1, except we now set $\psi=0$ and
find
\bea
\chi^\prime &=& \frac{3}{16}\, \epsilon^{4/3}\, h(\tau)\, K(\tau)^4\, \sinh^2\tau\, \lpl \sigma^\prime \,,
\\
\pr_\mu (\chi-\sigma) &+& \frac{9}{8}\, K(\tau)^2\,
\pr_\tau \Bigl\{ K^4\, \sinh^2\tau\, \pr_\mu \sigma^\prime\Bigr\} = 0\,.
\eea
Eliminating $\chi$ and changing variables as before,
\be
\tilde{w}={ \epsilon^{4/3} \over g_s M \alpha'} K^5\sinh(\tau)^2\sigma' \,,
\ee
we find
\be \label{RRscalar}
\tilde{w}''-{\cosh^2\tau+1\over \sinh^2 \tau}
\tilde{w}+{\epsilon^{4/3}h \over 6K^2}\lpl\tilde{w}=0 \,.
\ee
Again, after introducing the dimensionless mass as in (\ref{mass}), we get
a non-minimal scalar equation
\be
\label{RR_potential} \tilde{w}''-{\cosh^2\tau+1\over \sinh^2 \tau}
\tilde{w}+\mt^2 \frac{I(\tau)}{K(\tau)^2}\, \tilde{w}=0 \,.
\ee
Since the 4-d parity operation includes sign reversal of RR fields, we identify the
family of glueballs coming from this eigenvalue problem as {\it pseudoscalars} whose
$J^{PC}$ quantum numbers are $0^{--}$.

If we set $\tilde m=0$
the solution regular at small $\tau$ is $(\sinh 2\tau-2\tau)/\sinh \tau$.
Since this blows up at large $\tau$ we conclude that this equation does not contain
a massless glueball. A simple numerical analysis using the shooting method allows one to find the mass spectrum. The lowest eigenvalues ($\mt^2<100$) are listed in  Table~\ref{tab:eigenvals2}.
\begin{table}[tb]
\begin{center}
\caption{\small Non-zero eigenvalues with $\mt^2<100$ in the RR sector. This spectrum can also be fitted by a quadratic polynomial (red line).}
\label{tab:eigenvals2}
\begin{tabular}{cc}
\begin{tabular}[b]{||c|c||c|c||c|c||c|c||}
\hline
$n$ & $\mt_n^2$ & $n$ & $\mt_n^2$ & $n$ & $\mt_n^2$ & $n$ & $\mt_n^2$ \\
\hline
$1$ & 2.41 & $5$ & 14.0 & $9$ & 34.8 & $13$ & 64.7 \\
\hline
$2$ & 4.47 & $6$ & 18.0 & $10$ & 41.4 & $14$ & 73.7 \\
\hline
$3$ & 7.08 & $7$ & 23.2 & $11$ & 48.6 & $15$ & 83.2 \\
\hline
$4$ & 10.3 & $8$ & 28.7 & $12$ & 56.4 & $16$ & 93.3 \\
\hline
\end{tabular} &
\includegraphics[width=0.45\textwidth]{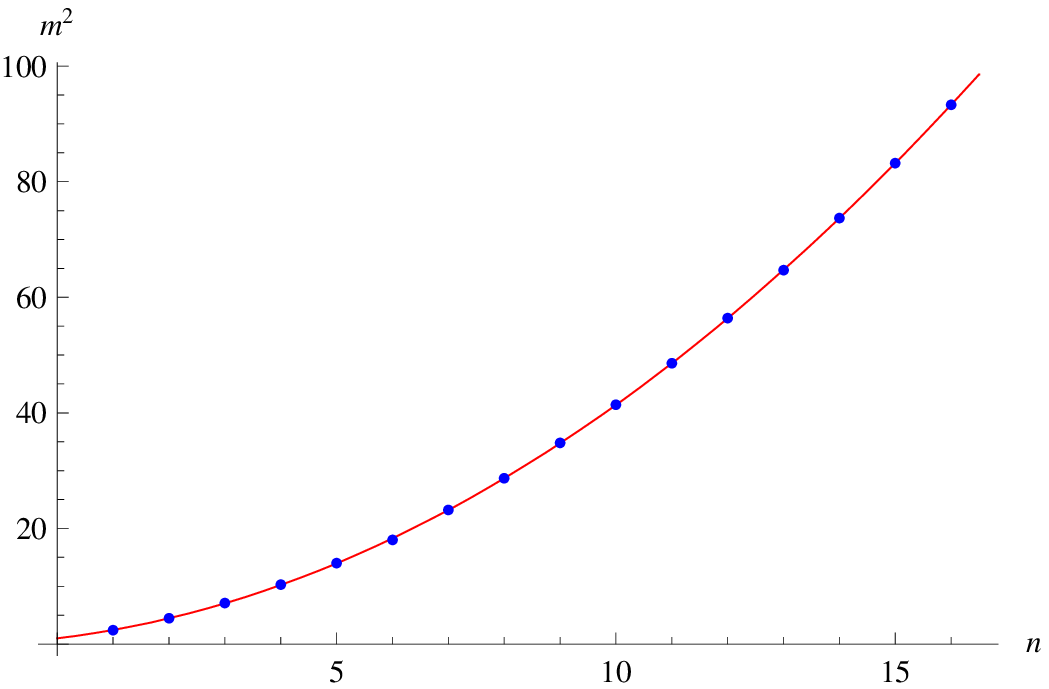}
\end{tabular}
\end{center}
\end{table}
The quadratic fit is
\be
\mt_{IIIn}^2 = 0.996 + 1.15\, n + 0.289\, n^2 \,;
\ee
in the BHM normalization it is given by
\be
m_{III\ BHM}^2 = 0.938 + 1.08\, n + 0.272\, n^2 \,.
\ee

The spectrum  can be reproduced with good accuracy using
a semiclassical (WKB) approximation. The effective potential in
(\ref{RR_potential}) is singular at $\tau=0$ which does not allow us
to use the conventional WKB approximation. Yet we can cast the
equation (\ref{RR_potential}) in the form $Q_1Q_2 \wt=m^2 \wt$,
where $Q_i$ are first-order differential operators and then
consider an equation $Q_2Q_1 \tilde{\wt}=m^2 \tilde{\wt}$, which
must give rise to the same spectrum up to a zero mode. Namely, in our case
this means that for $A$ such that
\bea
\label{A_eq} A^2+A'={\cosh^2\tau+1\over \sinh^2\tau}\ ,
\eea
equation (\ref{RR_potential}) shares the spectrum with an equation
\bea
\label{conjRR} \tilde{\tilde{w}}''-(B^2+B')\tilde{\tilde{w}}+\mt^2
\frac{I(\tau)}{K(\tau)^2}\, \tilde{\tilde{w}}=0 \, ,\\
B=-A-{1\over 2}{d\over d\tau}\log\frac{I(\tau)}{K(\tau)^2}\ .
\eea
A general solution of (\ref{A_eq}) reads
\bea
A=-\coth\tau+{2\sinh^2\tau\over
\cosh\tau\sinh\tau-\tau+\mathbf{C}}\ .
\eea
For (\ref{conjRR}) to be non-singular at the origin  $\mathbf{C}$
has to be non-zero. For a finite $\mathbf{C}$ the potential is
regular everywhere but not monotonic and (\ref{conjRR}) admits a
zero mode. A most convenient choice is to take infinite
$\mathbf{C}$, which reduces $A$ to $A=-\coth \tau$. In this case
the WKB approximation is applicable in it simplest form (see
\cite{Krasnitz} for similar considerations) and yields the same
result as the shooting method up to the third digit.

\section{Organizing the Modes into Supermultiplets}

\label{sec:superpartners}
The pseudoscalar Goldstone mode and the massless scalar found in \cite{GHK} belong to
a 4-dimensional chiral multiplet. These fields appear as the phase and the modulus of the
baryonic order parameters that vary along the baryonic branch.
When a long KS throat is embedded into a Calabi-Yau compactification with fluxes,
the baryonic $U(1)$ symmetry becomes gauged and a supersymmetric version of the Higgs mechanism
is expected to take place. The axial vector $U(1)_B$ gauge field `eats' the pseudoscalar mode and acquires a mass
degenerate with the mass of a scalar Higgs. These fields constitute the bosonic content of
a massive ${\cal N}=1$ axial vector supermultiplet.

In the present paper we explicitly constructed the massive modes that are radial excitations of the
GHK scalar. It is, of course, interesting to find the supermultiplets they belong to.
We will argue that each of these scalar radial excitations is also a member of a massive axial vector
supermultiplet. Similarly, each pseudoscalar glueball found in section~\ref{sec:another_scalar} is a member of
a massive vector multiplet. To prove these facts we would need to demonstrate the existence of
the $J^{PC}=1^{+-}$ glueballs degenerate with the $0^{+-}$ glueballs found in section~\ref{sec:radial_excitations}, as
well as of $1^{--}$ glueballs degenerate with the $0^{--}$ glueballs found in section~\ref{sec:another_scalar}.
Unfortunately, constructing decoupled equations for vector supergravity fluctuations around the KS
background is a difficult task. Instead, we will provide some evidence for
our claims
by studying axial vector and vector fluctuation equations in the large radius (KT) limit
(setting $\alpha' = g_s = 1$, $ N=0$ and $M = 2$; see Appendix~\ref{app:KT}).

First we reconsider the
simple decoupled pseudoscalar equation from the RR sector (\ref{RRscalar}) and argue that its
superpartner is given by the four-dimensional vector $A_1$ in
\be \label{VecA}
\delta B_2 = A_1 \we g^5 \ ,
\ee
where we have chosen the ansatz so that the corresponding radial component $ \sim dr \we g^5$ vanishes.
The equation for $d\ast H_3$ implies (with primes denoting derivatives with respect to $r$)
\be
\Bigl[{r^3 \over 6}\, \ast_4 A_1^\prime\Bigr]' - {4r \over 3}\, \ast_4 A_1 - {hr^3 \over 6}\, d_4 \ast_4 d_4\, A_1 = 0 \ ,
\ee
and $d_4 \ast_4 A_1 = 0$, i.e.\ the vector is divergence-free.
Since the Laplacian acting on such a vector is $\lpl = -\ast_4\, d_4\ast_4d_4$ (note the Minkowski signature of the four dimensional metric), we find
\be
\Bigl[{r^3 \over 6}\, A_1^\prime \Bigr]' - {4r \over 3}\, A_1 + {hr^3 \over 6}\, \lpl A_1 = 0 \ .
\ee
Defining a new variable $\tilde{A}_1 = r A_1$, it is easy to see that its equation of motion,
\be
{r\over 3} \Bigl[{r \over 3}\, \tilde{A}_1^\prime \Bigr]' - \tilde{ A_1} + {hr^2 \over 9}\, \lpl \tilde{A_1} = 0 \ ,
\ee
coincides with the KT-limit of the equation for the decoupled pseudoscalar
$\tilde{w}$, once we identify $r \sim \epsilon^{2/3} e^{\tau/3}$.
In fact, if we make the same ansatz (\ref{VecA}) in the full KS background,
 the equation of motion for $\tilde{A_1} = K^2 \sinh \tau A_1$ resulting from the terms $\sim d^3x \we d\tau \we \omega_2 \we \omega_2$ in $d\ast H_3 =0$ is precisely as in (\ref{RRscalar}):
\be
{d^2 \over d \tau^2}\, \tilde{A_1} - {\cosh^2\tau + 1 \over \sinh^2 \tau}\, \tilde{A_1} + {\epsilon^{4/3}h \over 6 K^2}\, \lpl\tilde{A_1} = 0 \ .
\ee
However, this ansatz is not closed in the KS case. The Bianchi identity for $F_5$ is not satisfied,
so this NSNS vector must mix with RR excitations of $F_3$ and/or $F_5$ in the KS background.
It would be interesting to solve this mixing problem.

Let us now turn to the massive axial vector superpartners of the coupled scalars
(\ref{CoupledScalarsz}), (\ref{CoupledScalarsw}) found above. We make the following ansatz, which is
similar to the one studied in \cite{MK},
\bea \label{vectoransatz}
\delta C_4 &=& B_1 \we \om_3 + F_2 \we \om_2 + K_1 \we dr \we \om_2\,, \\
\delta C_2 &=& C_1 \we g^5 + D_2 + E_1 \we dr\,, \\
\delta B_2 &=& H_2 + J_1 \we dr \,;
\eea
where $B_1, C_1, E_1, J_1, K_1$ are axial
vectors and $D_2, F_2, H_2$ are two-forms in four dimensions. We choose to split the six degrees of freedom residing in the two-form into a vector and a dual vector, e.g. $D_2 = d_4(\ldots) + \ast_4 d_4 D_1$. The degrees of freedom contained in the former (exact) part are in fact the same as those in $E_1$, so we can simply write
$D_2 = \ast_4 d_4 D_1$
without loss of generality. Similarly,
$F_2 = \ast_4 d_4 F_1$   and  $H_2 = \ast_4 d_4 H_1$,
and the corresponding exact parts can be absorbed into the vectors $K_1$ and $J_1$, respectively.

The equations of motion then imply that $B_1$ and $C_1$ have to be divergence-free: $d_4 \ast_4 B_1 = d_4 \ast_4 C_1 = 0$. If this were not the case their divergences would simply couple to additional scalars $\delta C_4 \sim dr \we \om_3$ and $\delta C_2 \sim dr \we g^5$, respectively, but we will not consider this here (i.e.~as for $A_1$ above we choose as gauge in which these radial components vanish).  In fact we will assume that all vectors in our ansatz are divergence-free, and that the terms appearing in the RR- and NSNS-potentials\footnote{I.e.~we demand that for example $\lpl B_1 = m^2 B_1$, but for the vectors derived from two-forms, such as $D_1$, we only impose the weaker condition $\lpl \ast_4 d_4 D_1 = m^2 \ast_4 d_4 D_1$.} are eigenstates of the Laplacian $\lpl = -\ast_4\, d_4\ast_4d_4$ with eigenvalue $m^2$.

We have relegated the details of the derivation of the equations of motion to Appendix~\ref{partners}.
Splitting the equations obtained from (\ref{FormEq}) into exact and coexact parts w.r.t.\ the four-dimensional derivative operator $d_4$ shows that the vectors $E_1$, $H_1$ and $K_1$ decouple\footnote{More precisely, we set ${ hr^5 \over 54 }\, E_1 = -3 \log{r \over r_\ast}\, H_1 = r \log{r \over r_\ast}\, K_1 $, and find a single second order differential equation obeyed by these fields. Thus we have found another decoupled vector, but this is not the one we are looking for. Given this relation between them, $E_1, H_1$ and $K_1$ do not mix with the other vectors.}. The resulting equations for the remaining vectors read
\be
\label{eq1v}
\Bigl[ \frac{3}{hr}\, B_1^\prime \Bigr]' + \frac{3}{r}\, \lpl B_1 = -\frac{3}{r}\, \lpl D_1 \,,
\ee
\be
\label{eq2v}
\Bigl[ \frac{r}{3}\, F_1^\prime \Bigr]' + \frac{hr}{3}\, \lpl F_1 = J_1 + \frac{3}{r}\, C_1 \,,
\ee
\be
\label{eq3v}
\Bigl[ \frac{r^3}{6}\, C_1^\prime \Bigr]' - \frac{4r}{3}\, C_1 + \frac{hr^3}{6}\, \lpl C_1 = \frac{3}{r} \lpl F_1 - \frac{9}{hr^2}\, B_1^\prime \,,
\ee
\be
\label{eq4v}
\Bigl[ \frac{hr^5}{54}\, D_1^\prime \Bigr]' + \frac{h^2 r^5}{54}\, \lpl D_1 = -F_1^\prime - \frac{3}{r}\, B_1 - 3\log\frac{r}{r_\ast} J_1 \,,
\ee
\be
\label{eq5v}
\Bigl[ \frac{hr^5}{54}\, J_1 \Bigr]' + 3\log\frac{r}{r_\ast} D_1^\prime = F_1^\prime + \frac{3}{r}\, B_1 \,,
\ee
\be
\label{eq6v}
\lpl F_1 - \frac{3}{hr}\, B_1^\prime = \frac{hr^5}{54}\, \lpl J_1 + 3 \log\frac{r}{r_\ast} \lpl D_1 \,,
\ee
where (\ref{eq2v}), (\ref{eq4v}) and (\ref{eq5v}) hold modulo terms annihilated by $d_4$.
It is easy to see that (\ref{eq5v}) and (\ref{eq6v}) imply (\ref{eq1v}), so the latter is not independent. We thus have the five coupled equations for the five vectors $B_1, C_1, D_1, F_1$ and $J_1$.

In the massless case, our ansatz includes the pseudoscalar found in \cite{GHK}. Putting
\bea
C_1 &=& - f_2(r) \ d_4 a(x)\ , \\
\lpl D_1 &=& f_1 \  d_4 a(x)\ , \\
B_1^\prime &=& - f_1 h r \log\frac{r}{r_\ast} \  d_4 a(x)\ , \\
F_1&=& J_1 = 0 \ ,
\eea
for some constant $f_1$ and a four-dimensional massless pseudoscalar $a(x)$, all equations of motion are satisfied provided
\be
\Bigl[ \frac{r^3}{6}\, f_2^\prime \Bigr]' - \frac{4r}{3} f_2 = - {9 \over r} \, f_1 \log\frac{r}{r_\ast}\ ;
\ee
in perfect agreement with the literature.

Now we would like to consider massive excitations however, and find axial
vector-like solutions to the equations (\ref{eq1v}) - (\ref{eq6v}) which give rise to the superpartners of the massive scalar excitations of (\ref{CoupledScalarsz}) and (\ref{CoupledScalarsw}). In particular, changing variables to $W_1 = r C_1$ equation (\ref{eq3v}) becomes
\be \label{Wpre}
{r \over 3}\,\left[{r \over 3}\, W_1^\prime \right]' - W_1 + {hr^2 \over 9}\, \lpl W_1 = {2 \over r} \lpl F_1 - {6 \over h r^2}\, B_1^\prime \ .
\ee
Thus we can identify $W_1$ with $\wt$ in (\ref{CoupledScalarsw}), which suggests setting the right hand side of this equation proportional to the counterpart of $\zt/r$. Hence we define
\be
Z_1 \equiv \lpl F_1 - \frac{3}{hr}\, B_1^\prime \,.
\ee
Using (\ref{eq1v}) and (\ref{eq2v}) one can deduce that this new field obeys
\be
{r \over 3}\,\left[{r \over 3}\, Z_1^\prime \right]'  + {hr^2\over 9}\, \lpl Z_1 = {1\over r}\, \lpl W_1+ {r\over 3}\, \lpl \bigl(J_1+D_1^\prime\bigr) \,.
\ee

Our reduced ansatz containing five axial
vectors is still too general. In order to match the spectrum of the scalar particles found above, we need to impose an additional  constraint to reduce the number of dynamical vectors obeying independent second order differential equations to two. The correct constraint for our purposes is given by
\be \label{constraint}
\lpl \bigl(J_1+D_1^\prime\bigr) = {3\over r^2}\, \lpl W_1.
\ee
In order to show that we can consistently impose this relation we need to examine the remaining equations. First of all, with this constraint (\ref{eq6v}) reads
\be
Z_1  = -\frac{hr^5}{54}\, \lpl D_1^\prime + \frac{hr^3}{18}\, \lpl W_1  + 3 \log\frac{r}{r_\ast} \lpl D_1\,.
\ee
Adding (\ref{eq4v}) and (\ref{eq5v}), using the constraint and the fact that $W_1$ is a mass eigenstate  we find
\be \label{WDeq}
\Bigl[ \frac{hr^3}{18}\, W_1 \Bigr]^\prime + {9 \over r^2}\log\frac{r}{r_\ast} W_1 + \frac{h^2 r^5}{54}\, \lpl D_1 =0 \,.
\ee
Eliminating $\lpl D_1$ between the last two equations we obtain a second order differential equation containing only $W_1$ and $Z_1$. A non-trivial fact is that this equation is identical to (\ref{Wpre}). This relies heavily on the precise expression for the warp factor (\ref{KTWarp}), and shows the consistency of the constraint equation with the equations of motion.

Finally, introducing a symbol for the other combination of the vectors $F_1$ and $B_1$ that appears in the equations of motion
\be
Y_1 \equiv F_1^\prime + {3\over r}\, B_1 \,,
\ee
equation (\ref{eq5v}) implies
\be \label{YZDeq}
\lpl Y_1 = Z_1^\prime - {3 \over r}\,  \lpl D_1.
\ee

In summary, we have the two coupled dynamical equations
\bea \label{Zeq}
{r \over 3}\,\left[{r \over 3}\, Z_1^\prime \right]'  + {hr^2\over 9}\, \lpl Z_1 &=& {2\over r}\, \lpl W_1 \ , \\
\label{Weq}
{r \over 3}\,\left[{r \over 3}\, W_1^\prime \right]' - W_1 + {hr^2 \over 9}\, \lpl W_1 &=& {2 \over r} Z_1 \ ,
\eea
which determine $W_1$ and $Z_1$. In terms of these $\lpl D_1$ is determined by (\ref{WDeq}), $J_1$ by (\ref{constraint}), and $\lpl Y_1$ by  (\ref{YZDeq}). Equations (\ref{Zeq}), (\ref{Weq}) are precisely the KT limit of
the scalar equations (\ref{CoupledScalarsz}), (\ref{CoupledScalarsw})
up to a rescaling of the fields by a numerical factor.\footnote{
 Looking at (\ref{CoupledScalarsz}) one might  have expected the term $2 \tilde{z}/\sinh^2 \tau$
to give rise to a term proportional to $Z_1/r^6$ in  (\ref{Zeq}), but in fact this is not
the case because it is too small to be seen in the KT limit.
In the KS background it arises from a subleading term in the variation of the
Ricci tensor (\ref{Ricci}) but such terms that are asymptotically suppressed by
powers of $r$ compared to the leading terms are not taken into account in the KT metric.
Indeed, if we write ansatz (\ref{ansatz}) in the KT background and follow the same strategy as we did for the full KS background,
the term proportional to $\zt/r^6$ does not appear in the Einstein equations.}

Since the KT limits of their equations of motion agree, we thus argue that the axial
vectors $Z_1$ and $W_1$ are the superpartners of the coupled scalars $\tilde{z}$ and $\tilde{w}$ found
above, and their massive excitations combine into vector multiplets.

\section{Effects of Compactification}
\label{sec:perturbative}
Now we will embed the KS throat into
a flux compactification, along the lines of \cite{Giddings},
and estimate the mass of the
Higgs scalar. Generally, glueballs are dual to the normalizable modes localized near the bottom of
the throat, and one does not expect them to be strongly
affected by the bulk of the Calabi-Yau. This is indeed the case for all the massive radial excitations found in sections~\ref{sec:radial_excitations} and \ref{sec:another_scalar}. We will see, however, that the case of the
GHK scalar is more subtle  and exhibits some UV sensitivity.

To model a compactification, we will
introduce a UV cut-off on the radial coordinate, $\tau_{\max}$.
We also need to include a deformation of
the KS solution introduced by bulk effects. On the field theory side this corresponds
to perturbing the Lagrangian of the cascading gauge theory by some irrelevant operators.
Here we are not interested in classifying all of them but
rather model the compactification effects in the simplest way by
considering one perturbation which simulates the main features of
the compactified solution. We consider a shift of the warp
factor $\delta h={\rm const}$ which corresponds to the dimension
$8$ operator on the field theory side \cite{GHKK,Kachru}.
This also has a simple geometrical meaning: the warp factor of
the compactified solution is a finite constant in the bulk of the
Calabi-Yau and therefore should not drop below a certain value along
the throat.

Let us introduce a small parameter $\delta$ which shifts the rescaled warp
factor, $I(\tau)\to I(\tau) + \delta$, and consider the system
(\ref{eq1f})-(\ref{eq2f}) in perturbation theory near $\tilde{m}^2=0$:
\bea
\zt=\zt_0+ \tilde{m}^2  \zt_1\ ,\\
\wt=\wt_0+ \tilde{m}^2  \wt_1\ , \\
\zt_0=\tau\coth\tau -1 \ ,\\
 \wt_0(\tau) = -\, \frac{2^{2/3}\,
8}{9}\, \frac{1}{\sinh\tau} \int_0^\tau\! dx\, I(x) \sinh^2 x \,.
\eea
At
leading order in $\tilde{m}^2$
\bea
\zt_{1} &=& (\tau\coth\tau-1) \int_0^\tau\! dx\, u(x) \coth x -
\coth \tau \int_0^\tau\! dx\, u(x)\, (x \coth x -1) \,,
\\
\wt_{1} &=& - \frac{1}{4\, \sinh\tau} \int_0^\tau\! dx\, v(x)\,
\frac{\sinh 2x-2x}{\sinh x} - \frac{\sinh 2\tau-2\tau}{4\,
\sinh\tau} \int_\tau^\infty\! dx\, v(x)\, \frac{1}{\sinh  x} \,.
\eea
\bea
u(\tau) &=& -{I(\tau)\over K^2(\tau)}\, \zt_0 + {9 \over 4\cdot
2^{2/3}}\, K(\tau)\, \wt_0  -\frac{\delta}{K^2}\, \zt_0\, ,\\
v(\tau) &=& -{I(\tau)\over K^2(\tau)}\, \wt_0 + {16\over 9}\,
K(\tau)\, \zt_{1} -\frac{\delta}{K^2}\, \wt_0 \ .
\eea
Keeping in mind that for large $\tau$, $u\simeq -2^{-2/3} \delta
\tau \e^{2\tau/3}$ one finds the asymptotic behavior
\bea
\zt_1(\tau) &\simeq& -2^{-2/3} \delta \int_0^\tau\! dx\, (\tau-x) x \e^{2x/3} \;\simeq\; -\frac{9\,\delta}{4\, 2^{2/3}}\, \tau \e^{2\tau/3} \,.
\eea
This yields $v \simeq - 2^{2/3} \delta \tau \e^{\tau/3}$ and
\bea
\wt_1 &=& - \frac{1}{4\, \sinh\tau} \int_0^\tau\! dx\, v_0(x)\, \frac{\sinh 2x-2x}{\sinh x} - \frac{\sinh 2\tau-2\tau}{4\, \sinh\tau} \int_\tau^\infty\! dx\, v_0(x)\, \frac{1}{\sinh  x} \;\simeq\; \frac{9\, 2^{2/3} \delta}{8}\, \tau \e^{\tau/3} \,.\qquad
\eea
Finally, up the first order in the mass squared and $\delta$:
\bea
\zt\;\simeq\; \tau\, \bigl[1-\frac{9\,\delta\mt^2}{4\, 2^{2/3}}\, \e^{2\tau/3} \bigr] \,, &\qquad & \wt \;\simeq\; -2^{4/3} \tau \e^{-\tau/3}\, \bigl[1-\frac{9\,\delta\mt^2}{8\, 2^{2/3}}\, \e^{2\tau/3} \bigr] \,.
\eea
This suggests that for generic boundary conditions the cut-off value
\be
\label{t_cutoff} \tau_{\max} \simeq -\log{\delta^{3/2} \tilde m^3} \,.
\ee
\begin{figure}[tb]
\begin{center}
\includegraphics[width=0.70\textwidth]{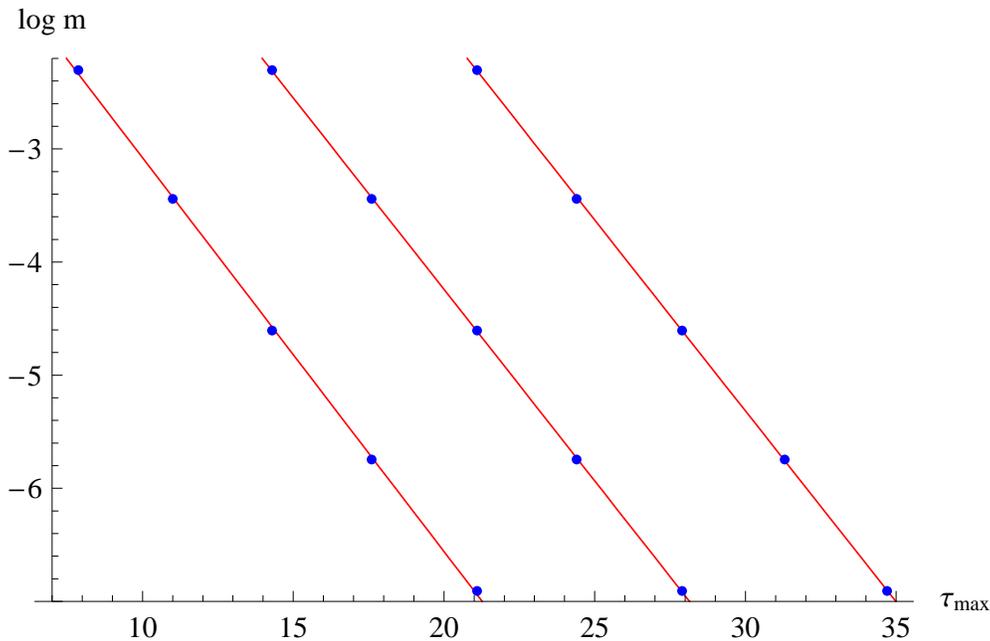}
\end{center}
\caption{\small The dependence of $\log \tilde m$ on $\tau_{\max}$ is
linear with the slope equal to -$1/3$. The three lines shown correspond to $\delta=1$, $\delta=0.01$ and $\delta=0.0001$.}
\label{fig:tcutoff}
\end{figure}
This prediction can be tested numerically.  In  order to do this
one can specify some small $\tilde m$ and plot the determinant
\bea
\label{rkmat} \text{det}\, \left(\ba[ccc]
\zt_1(\tau) &\; & \zt_2(\tau) \\
\wt_1(\tau) &\; & \wt_2(\tau) \ea\right) \,,
\eea
of the two
linearly independent solutions regular at $\tau=0$ as a function
of $\tau$. The first zero marks the point $\tau_{\max}$ such that
there is a regular solution with $z(\tau_{\max})=w(\tau_{\max})=0$.
Hence $\tau_{\max}$ is the corresponding cut-off value.
As Fig.~\ref{fig:tcutoff} shows, the relation (\ref{t_cutoff}) holds
for $\tau_{\max}$ large enough that
\be \label{smallmass}
\tilde m^2\sim \delta^{-1} e^{-2\tau_{\max}/3}
\ee
is small.

Let us consider a simple model of compactification where the throat is embedded
into an asymptotically conical space that terminates at some large
cut-off value $\tau_{\max}$.
To calculate the mass from (\ref{smallmass}) we need to know
$\delta$ as well as $\tau_{\max}$. The former is the asymptotic
value of the (rescaled) warp factor. The point where the field theory warp
factor approaches $\delta$ marks the UV cutoff of the field
theory
\bea
I(\tau_{UV})\sim \tau_{UV}e^{-4\tau_{UV}/3}\simeq \delta\ .
\eea
Using this in (\ref{smallmass}) we find
$\tilde m^2\sim e^{(4\tau_{UV}-2\tau_{\max})/3}$. This shows that the Higgs mass becomes
parametrically small only for $\tau_{\max}\gg 2\tau_{UV}$. This is not satisfied in general; the
geometry requires only that $\tau_{\max}> \tau_{UV}$
because $\tau_{UV}$ is the length of the throat embedded into a CY space.
With the ratio between the UV and IR scales of the field theory
around $4\cdot 10^3$ \cite{KKLMMT} we estimate that
$\tau_{UV}\simeq 25$ \cite{DKS}.
The cut-off
$\tau_{\max}$ can be related to the warped volume of the
Calabi-Yau
which, in a singular conifold approximation, is
\bea
V_6^w={\rm Vol}(T^{1,1})\int_0^{r_{\max}} dr h(r) \sqrt{{\rm det}\
g_6\over {\rm det}\  g_{T^{1,1}}} \ ,
\eea
where
$r\sim \epsilon^{2/3} e^{\tau/3}$.
The integral from zero to $r_{UV}$ is the warped volume of the throat, and
from $r_{UV}$ to $r_{\max}$ is the bulk volume. Assuming that the latter dominates,
\bea
V_6^w \simeq {16\pi^3\over 27}\epsilon^{4/3}
(g_sM\alpha')^2\left[r^6_{\max}-r^6_{UV}\right]r^{-4}_{UV}
\ .
\eea
Requiring $\tau_{\max}\gg 50$ leads to an enormous $V_6^w$, far larger than, for example,
$V_6^w\simeq 5^6 \alpha'^3$ in \cite{KKLMMT}.

Thus, while for $\tau_{\max}\gg 2\tau_{UV}$ the Higgs scalar becomes parametrically
lighter than the other normal modes, in compactifications with realistic parameters it may actually
be heavier. This is due to the special feature of its wave function $\tilde z$ which
grows linearly with $\tau$ in the throat.
The only conclusion we can draw from our simplified model of compactification is that
this mode is rather UV sensitive, so to determine its mass
we need to know the details of the compactification.

\section{Acknowledgements}

We thank S. Giddings,
C. Herzog, J. Maldacena and M. Reece for useful discussions. This research was supported
in part by the National Science Foundation under Grant No.
PHY-0243680. The research of AD is also supported in part by Grant
RFBR 07-02-00878, and Grant for Support of Scientific Schools
NSh-8004.2006.2.  Any opinions, findings, and conclusions or
recommendations expressed in this material are those of the
authors and do not necessarily reflect the views of the National
Science Foundation.

\appendix

\section{The Type IIB Supergravity Equations}
\label{app:SUGRA_eqns}
Here we succinctly list the equations of motion required to study RR and NSNS
2-form perturbations. Since the dilaton and RR scalar do not enter at linear
order, we set them to zero.

Bianchi identities:
\be\ba \label{BianchiId}
dF_3 &=& 0 \,,
\\
dH_3 &=& 0 \,,
\\
dF_5 &=& H_3 \wedge F_3 \,.
\ea\ee

Dynamic equations:
\be\ba \label{FormEq}
d \star H_3 &=& - g_s^2 F_5 \wedge F_3 \,,
\\
d \star F_3 &=& F_5 \wedge H_3 \,,
\\
 F_5 &=& \star F_5 \,.
\ea\ee

Einstein equation:
\be \label{lEE}
R_{ij}  = T_{ij} =  {g_s^2\over 96} F_{iabcd}F_j^{\ abcd}+
{1\over 4}H_{iab} H_j^{\ ab} -{1\over 48}G_{ij} H_{abc} H^{abc}
 + {g_s^2\over 4}F_{iab} F_j^{\ ab} -{g_s^2\over 48}G_{ij} F_{abc} F^{abc}
\ .
\ee

\section{Review of Warped Deformed Conifolds}
\subsection{The KS Solution}
\label{app:KS_solution}
The ten dimensional metric for the KS solution  is
\be \label{10dmetric}
ds_{10}^2 = h(\tau)^{-1/2} (-dt^2 + dx^2 + dy^2+dz^2)
+h(\tau)^{1/2} ds_6^2 \ ,
\ee
where
\be \label{conifoldmetric}
ds_6^2 = {\epsilon^{4/3} K(\tau) \over 2}
\left[ {1 \over 3K^3} (d\tau^2 + (g_5)^2)
 + \cosh^2 \left ({\tau\over 2}\right ) ((g^3)^2 + (g^4)^2)
+ \sinh^2 \left ({\tau\over 2}\right ) ((g^1)^2 + (g^2)^2) \right]
\ee
is the usual warped deformed conifold metric.
The volume form is
\begin{equation}
\vol = {\epsilon^4\over 96} h^{1/2} \sinh^2\tau
dt \wedge dx^1 \wedge dx^2 \wedge dx^3\wedge d\tau \wedge g^1\wedge g^2
\wedge g^3\wedge g^4\wedge g^5 \,.
\end{equation}
The one-forms are given in terms of angular coordinates as
\begin{eqnarray} \label{gbasis}
g^1 = {e^1-e^3\over\sqrt 2}\ ,\qquad
g^2 = {e^2-e^4\over\sqrt 2}\ , \nonumber \\
g^3 = {e^1+e^3\over\sqrt 2}\ ,\qquad
g^4 = {e^2+ e^4\over\sqrt 2}\ , \nonumber \\
g^5 = e^5\ ,
\end{eqnarray}
where
\begin{eqnarray} \label{ebasis}
e^1\equiv - \sin\theta_1 d\phi_1 \ ,\qquad
e^2\equiv d\theta_1\ , \nonumber \\
e^3\equiv \cos\psi\sin\theta_2 d\phi_2-\sin\psi d\theta_2\ , \nonumber\\
e^4\equiv \sin\psi\sin\theta_2 d\phi_2+\cos\psi d\theta_2\ , \nonumber \\
e^5\equiv d\psi + \cos\theta_1 d\phi_1+ \cos\theta_2 d\phi_2 \ .
\end{eqnarray}
Also
\bea
de^1 &=& - \frac{\cos\theta_1}{\sin\theta_1}\, e^1 \wg e^2 \,,
\\
de^2 &=& 0 \,,
\\
de^3 &=& e^4 \wg e^5 + \frac{\cos\theta_1}{\sin\theta_1}\, e^4 \wg e^1 \,,
\\
de^4 &=& -e^3 \wg e^5 - \frac{\cos\theta_1}{\sin\theta_1}\, e^3 \wg e^1 \,,
\\
de^5 &=& -e^1 \wg e^2 + e^3 \wg e^4
\eea
and
\bea
dg^1 &=& \half\, (g^2-g^4)\wg g^5 -\frac{1}{\sqrt{2}}\, \cot\theta_1\, (g^1+g^3)\wg g^2 \,,
\\
dg^2 &=& - \half\, (g^1-g^3)\wg g^5 - \frac{1}{\sqrt{2}}\, \cot\theta_1\, g^1\wg g^3 \,,
\\
dg^3 &=& -\half\, (g^2-g^4)\wg g^5 -\frac{1}{\sqrt{2}}\, \cot\theta_1\, (g^1+g^3)\wg g^4 \,,
\\
dg^4 &=& -dg^2 \,,
\\
dg^5 &=& - (g^1\wg g^4 + g^3\wg g^2) \,.
\eea
Note that
\be \label{Keqn}
K(\tau)= { (\sinh (2\tau) - 2\tau)^{1/3}\over 2^{1/3} \sinh \tau}
\ .
\ee
The warp factor is
\be \label{intsol}
h(\tau) = (g_s M\alpha')^2 2^{2/3} \varepsilon^{-8/3} I(\tau)\ ,
\ee
where
\be \label{Itau}
I(\tau) \equiv
\int_\tau^\infty d x {x\coth x-1\over \sinh^2 x} (\sinh (2x) - 2x)^{1/3}
\ .
\ee
The NSNS two-form field and corresponding field strength are
\be \label{B2}
B_2 = {g_s M \alpha'\over 2} [f(\tau) g^1\wedge g^2
+  k(\tau) g^3\wedge g^4 ]\ ,
\ee
\begin{eqnarray}
\label{H3}
H_3 = dB_2 &=& {g_s M \alpha'\over 2} \left[
d\tau\wedge (f' g^1\wedge g^2
+  k' g^3\wedge g^4) + {1\over 2} (k-f)
g^5\wedge (g^1\wedge g^3 + g^2\wedge g^4) \right]\ ,
\end{eqnarray}
while the RR three-form field strength is
\begin{eqnarray}
\label{F3}
F_3 &=& {M\alpha'\over 2} \Bigl [g^5\wedge g^3\wedge g^4 + d [ F(\tau)
(g^1\wedge g^3 + g^2\wedge g^4)]\Bigr ]
\\  \nonumber
&=&  {M\alpha'\over 2} \Bigl[ g^5\wedge g^3\wedge g^4 (1- F)
+ g^5\wedge g^1\wedge g^2 F + F' d\tau\wedge
(g^1\wedge g^3 + g^2\wedge g^4) \Bigl]\ .
\end{eqnarray}
The auxiliary functions in these forms are
\begin{eqnarray}
F(\tau) &=& {\sinh \tau -\tau\over 2\sinh\tau}\ ,
\nonumber \\
f(\tau) &=& {\tau\coth\tau - 1\over 2\sinh\tau}(\cosh\tau-1) \ ,
\\ \nonumber
k(\tau) &=& {\tau\coth\tau - 1\over 2\sinh\tau}(\cosh\tau+1)
\ .
\end{eqnarray}
Some useful identities are
\be\ba
k-f &=& 2\, F^\prime \,,
\\
f^\prime &=& (1-F)\, \tanh^2(\tau/2) \,,
\\
k^\prime &=& F\, \coth^2(\tau/2) \,.
\ea\ee
The five-form field strength is given by
\be
F_5 =(1+\ast)\, B_2\wedge F_3\ .
\ee
We also note that
\be \label{dg5}
dg^5 = - (g^1 \wg g^4 + g^3\wg g^2) \ ,
\ee
and
\be
dg^5 \wg dg^5 = - 2 g^1 \wg g^2 \wg g^3 \wg g^4 \ .
\ee

\subsection{The KT Solution}
\label{app:KT}
The KT solution~\cite{KT} corresponds to the large $\tau$ limit of the more general KS solution.
For simplicity we take $g_s = \alpha' = 1$, $M=2$ and $N=0$. In terms of the radial coordinate
$r \sim \epsilon^{2/3} e^{\tau/3}$ the KT background is given by
\bea
ds^2 &=& \frac{1}{\sqrt{h}}\, (-dt^2+dx^2) + \sqrt{h} (dr^2 + r^2 ds_{T^{11}}^2) \,,
\\
H_3 &=& \frac{3}{r}\, dr\wedge \omega_2 \,, \qquad B_2 \;=\; 3\, \log\frac{r}{r_\ast}\, \omega_2 \,,
\qquad F_3 =  \omega_3 \,,
\\
F_5 &=& (1+\ast)\, B_2\wedge F_3 \;=\; 3 \log\frac{r}{r_\ast}\, \Bigl[ \omega_2 \wedge \omega_3 - \frac{54}{h^2 r^5} \, d^4 x \wedge dr \Bigr] \,.
\eea
The warp factor is given by
\bea
\label{KTWarp}
h(r) = {81 \over 8 r^4} \left(1+4 \log{r \over r_\ast}\right)\ ,
\eea
and the conifold metric is
\bea
ds_{T^{11}}^2 &=& \frac{1}{9} (g^5)^2 + \frac{1}{6} \sum_{i=1}^4 (g^i)^2 \, .
\eea
The volume form is given by
\bea \label{KTVol}
\vol &=& \frac{\sqrt{h}r^5}{54}\, d^4x \wedge \omega_2 \wedge \omega_3 \wedge dr \,.
\eea
Here we have introduced the two harmonic forms,
\be \label{defo2}
\omega_2 = \frac{1}{2}(g^1 \wedge g^2 + g^3 \wedge g^4) = \frac{1}{2}
(\sin\theta_1 d\theta_1 \wedge d\phi_1 - \sin \theta_2 d\theta_2 \wedge d\phi_2) \ ,
\ee
\be \label{defo3}
\omega_3 = \omega_2 \we g^5 \ .
\ee

\subsection{BHM Normalization}
\label{app:BHM}
Here we show how to find the conversion factor between the dimensionless mass squared $\mt^2$ and the mass in the normalization of Berg, Haack and M\"uck~\cite{BHM2}. We note that the BHM conventions correspond to the KS solution with an extra relation between $\epsilon$ and $M$. The authors of~\cite{BHM2} use the notations of the general PT ansatz (as given in Eq.~(3.8) of \cite{BHM1}):
\bea
ds^2 &=& \e^{2p-x} ds_5^2 + (\e^{x+g}+a^2\e^{x-g})(e_1^2+e_2^2) + \e^{x-g}[e_3^2+e_4^2-2a(e_1e_3+e_2e_4)] + \e^{-6p-x}e_5^2 \,,
\\
ds_5^2 &=& dr^2 + \e^{2A(r)} \eta_{ij} dx^i dx^j \,.
\eea
After setting\footnote{The Papadopoulos-Tseytlin
\cite{PT} variables are $(x,p,y,\Phi,b,h_1,h_2)$.}
\be
a=\tanh y=\frac{1}{\cosh\tau}\,, \quad \e^{-g}=\cosh y=\coth\tau \,;
\ee
it reduces to the KS form
\be
ds^2 = \e^{2A+2p-x}\eta_{ij}dx^i dx^j + \frac{\e^x}{\sinh\tau} \Bigl[ \coth\tau (e_1^2+e_2^2+e_3^2+e_4^2) + \frac{2}{\sinh\tau} (e_1e_3+e_2e_4) + \e^{-6p-2x} (d\tau^2+e_5^2) \Bigr] \,.
\ee
The radial KS coordinate $\tau$ is introduced according to
\be
\pr_\tau = \e^{-4p} \pr_r \,.
\ee
Note that in the KS notation the conifold metric (\ref{conifoldmetric}) can be rewritten as
\be
ds_6^2 = {\epsilon^{4/3} K(\tau) \over 2}
\left[ {1 \over 3K^3} (d\tau^2 + (e_5)^2)
+ \half\cosh\tau(e_1^2 + \e_2^2 + e_3^2 + e_4^2) + e_1e_3+e_2e_4 \right] \,.
\ee

In terms of $\tau$, the PT variables necessary to describe the metric for the KS background solution take the form
\bea
\label{KS:Phisol}
\Phi &=& \Phi_0\,,\\
\label{KS:ysol}
\e^{y} &=& \tanh (\tau/2)\,, \\
\label{KS:xsol}
\frac23\, \e^{6p+2x} &=& \coth \tau -\frac{\tau}{\sinh^2 \tau}\,,\\
\label{KS:psol}
\e^{2x/3-4p} &=& 6^{-2/3} M^2 \e^{\Phi_0} I(\tau) \sinh^{4/3} \tau\,.
\eea
In the BHM normalization
\begin{equation}
\e^{-2A-8p} = \left( \e^{-6p-2x} \sinh\tau \right)^{2/3} \frac{I(\tau)}{I_0} \,, \qquad I_0 \equiv I(0) \,.
\end{equation}
These equations give for the coefficients
\bea
\e^{6p+2x} &=& \frac{3}{2}\, K^3\sinh\tau \,,
\\
\e^x &=& 2^{-2/3} \e^{\Phi_0/2} MK(\tau)\sinh\tau \sqrt{I(\tau)} \,,
\\
\e^{2A+2p-x} &=& \sqrt{\e^{2x/3-4p}} \sinh^{-2\tau/3} \frac{I_0}{I} \;=\; 6^{-1/3} \e^{\Phi_0/2} M \frac{I_0}{\sqrt{I}} \,.
\eea
Comparing these coefficients with those of the KS solution we find\footnote{We set $g_s=\alpha^\prime=1$ according to~\cite{BHM2}.}
\bea
\frac{\epsilon^{4/3}}{M^2} &=& 3^{-1/3} \e^{\Phi_0/2} I_0 \,,
\\
\e^{\Phi_0/2} &=& \half \,.
\eea

This yields $\epsilon^{4/3}/M^2=3^{-1/3}I_0/2$.
Then using (\ref{mass}) we get for the four-dimensional mass in the BHM normalization
\be
m_{BHM}^2 = m_4^2 = \mt^2 \, \frac{6}{2^{2/3}}\, \frac{\epsilon^{4/3}}{M^2} = (3/2)^{2/3} I_0\, \mt^2 \,.
\ee

\section{Equations of Motion for Vector Superpartners}
\label{partners}
With the ansatz (\ref{vectoransatz}), the deformations of the field strengths are
\bea
\delta H_3 &=& - \ast_4 \lpl H_1 + \bigl( \ast_4 d_4 H_1^\prime + d_4 J_1 \bigr) \wedge dr \,,
\\
\ast \delta H_3 &=& - \frac{h^2 r^5}{54}\, \lpl H_1 \wedge \omega_2 \wedge \omega_3 \wedge dr + \frac{h r^5}{54}\, \bigl( d_4 H_1^\prime - \ast_4 d_4 J_1 \bigr) \wedge \omega_2 \wedge \omega_3 \,;
\eea
\bea
\delta F_3 &=& d_4 C_1 \wedge g^5 - C_1^\prime \wedge dr \wedge g^5 - C_1 \wedge dg^5 + \bigl( \ast_4 d_4 D_1^\prime + d_4 E_1 \bigr) \wedge dr + d_4 \ast_4 d_4 D_1 \,,
\\
\nn \ast \delta F_3 &=& \frac{h r^3}{6}\, \ast_4 d_4 C_1 \wedge \omega_2 \wedge \omega_2 \wedge dr + \frac{r^3}{6}\, \ast_4 C_1^\prime \wedge \omega_2 \wedge \omega_2  + \frac{r}{3}\, \ast_4 C_1 \wedge dg^5 \wedge g^5 \wedge dr
\\
&& + \frac{hr^5}{54}\, \bigl( d_4 D_1^\prime - \ast_4 d_4 E_1 \bigr) \wedge \omega_2 \wedge \omega_3 - \frac{h^2 r^5}{54}\,  \lpl D_1 \wedge \omega_2 \wedge \omega_3 \wedge dr \,;
\eea
\bea
\delta F_5 &=& \delta \cF_5 + \ast \delta \cF_5 \,,
\\
\delta \cF_5 &=& \bigl( d_4 B_1 - B_1^\prime \wedge dr \bigr) \wedge \omega_3 + \bigl( - \ast_4 \lpl F_1 + \ast_4 d_4 F_1^\prime \wedge dr \bigr) \wedge \omega_2 + d_4 K_1 \wedge dr \wedge \omega_2 \,,
\\
\nn \ast \delta \cF_5 &=& \frac{3}{r}\, \ast_4 d_4 B_1 \wedge \omega_2 \wedge dr + \frac{3}{hr}\, \ast_4 B_1^\prime \wedge \omega_2 - \frac{hr}{3}\, \lpl F_1 \wedge \omega_3 \wedge dr
\\
&& + \frac{r}{3}\, \bigl( d_4 F_1^\prime - \ast_4 d_4 K_1 \bigr) \wedge \omega_3 \,.
\eea
The equations of motion that result from this ansatz are as follows. The Bianchi identity for $F_5$ gives
\bea
\frac{r}{3}\, \ast_4 \lpl K_1&=& -\ast_4 \lpl H_1 \,,
\\
- \frac{3}{r}\, \ast_4 \lpl B_1 - \Bigl[ \frac{3}{hr} \ast_4 B_1^\prime \Bigr]' &=& \frac{3}{r}\, \ast_4 \lpl D_1 \,,
\\
\frac{hr}{3}\, d_4 \lpl F_1 + \Bigl[ \frac{r}{3} \bigl( d_4 F_1^\prime - \ast_4 d_4 K_1 \bigr) \Bigr]' &=& \ast_4 d_4 H_1^\prime + d_4 J_1 + \frac{3}{r}\, d_4 C_1 \,.
\eea
From the equations for $\ast F_3$ in (\ref{FormEq}) we have
\bea
\frac{hr^5}{54}\, \ast_4 \lpl E_1 &=& -3 \log\frac{r}{r_\ast} \ast_4 \lpl H_1 \,,
\\
- \frac{hr^3}{6} \ast_4 \lpl C_1 - \Bigl[ \frac{r^3}{6} \ast_4 C_1^\prime \Bigr]' + \frac{4r}{3} \ast_4 C_1 &=& -\frac{3}{r} \ast_4 \lpl F_1 + \frac{9}{hr^2} \ast_4 B_1^\prime \,,
\\
\nn \Bigl[ \frac{hr^5}{54} \bigl( d_4 D_1^\prime - \ast_4 d_4 E_1 \bigr) \Bigr]' + \frac{h^2 r^5}{54}\, d_4 \lpl D_1 &=& \ast_4 d_4 K_1 - d_4 F_1^\prime
\\
&& - 3\log\frac{r}{r_\ast} \bigl( \ast_4 d_4 H_1^\prime + d_4 J_1 \bigr) - \frac{3}{r}\, d_4 B_1 \,,
\eea
and from the equations for $\ast H_3$
\bea
\frac{hr^5}{54} \ast_4 \lpl J_1 &=& -3 \log\frac{r}{r_\ast} \ast_4 \lpl D_1 + \ast_4 \lpl F_1 - \frac{3}{hr} \ast_4 B_1^\prime \,,
\\
\nn \frac{h^2 r^5}{54} d_4 \lpl H_1 + \Bigl[ \frac{hr^5}{54} \bigl( d_4 H_1^\prime - \ast_4 d_4 J_1\bigr) \Bigr]' &=& 3 \log\frac{r}{r_\ast} \bigl( \ast_4 d_4 D_1^\prime + d_4 E_1 \bigr)
\\
&& - \ast_4 d_4 F_1^\prime - d_4 K_1 - \frac{3}{r} \ast_4 d_4 B_1 \,.
\eea

\section{Numerical Analysis: Finding the Spectra}
\label{app:numerics} 
A standard method of finding the
spectrum of a single second-order differential equation is the
shooting technique. For a system of several coupled linear equations
the shooting method has to be generalized \cite{BHM2}. Here we will focus on the
subtleties specific to the system of equations (\ref{eq1f}) and
(\ref{eq2f}). The idea of the calculation (called the determinant
method \cite{BHM2}) is to set the initial conditions at infinity corresponding
to the two solutions regular at infinity, $\left(\ba[c] \zt_1(\tau) \\
\wt_1(\tau) \ea\right)$ and $\left(\ba[c] \zt_2(\tau) \\
\wt_2(\tau) \ea\right)$, and extend them numerically to small
$\tau$. Then the matrix
\be
\label{rkmatnew} \left(\ba[ccc]
\zt_1(0) &\; & \zt_2(0) \\
\wt_1(0) &\; & \wt_2(0)
\ea\right)
\ee
becomes degenerate at the critical points (eigenvalues) in the
spectral parameter space.

Let us find the asymptotic behavior of regular and singular
solutions near both zero and infinity. At small $\tau$ equations
(\ref{eq1f}) and (\ref{eq2f}) decouple,
\bea
\zt'' - \frac{2}{\tau^2}\, \zt &=& 0 \,,
\\
\wt'' - \frac{2}{\tau^2}\, \wt &=& 0 \,.
\eea
There are the two regular solutions with $\zt,\: \wt\sim \tau^2$ and
the two singular solutions with $\zt,\: \wt\sim 1/\tau$. For large
$\tau$ we have
\bea
\zt'' &=& \mt^2 \frac{9}{4\cdot 2^{1/3}}\, \e^{-\tau/3} \wt \,,
\\
\wt''-\wt &=& \frac{16\cdot 2^{1/3}}{9}\, \e^{-\tau/3} \zt \,.
\eea
The asymptotic behavior of the two regular solutions is
\bea
\left(\ba[c] \zt_1\\ \wt_1\ea\right) \;=\; \left(\ba[c] 1\\
-2^{4/3}\,
\e^{-\tau/3} \ea\right) \,, &\qquad& \left(\ba[c] \zt_2\\
\wt_2\ea\right) \;=\; \left(\ba[c] \frac{81}{64\cdot 2^{1/3}}\,
\mt^2 \e^{-4\tau/3}\\ \e^{-\tau} \ea\right) \,;
\eea
and the singular solutions are
\bea
\left(\ba[c] \zt_3\\ \wt_3\ea\right) \;=\; \left(\ba[c] \tau\\
-2^{4/3}\, \bigl(\tau-\frac{3}{4}\bigl) \e^{-\tau/3} \ea\right) \,,
&\qquad& \left(\ba[c] \zt_4\\ \wt_4\ea\right) \;=\; \left(\ba[c]
\frac{81}{16\cdot 2^{1/3}}\, \mt^2 \e^{2\tau/3}\\ \e^\tau\ea\right)
\,.
\eea

A particular subtlety  of this setup is that at large $\tau$ the
two singular solutions don't diverge equally fast: one of them grows
exponentially while the other is only  linear in $\tau$. This makes
it difficult to start shooting from zero: imposing the regularity
condition at infinity would require vanishing of both linear and
exponential terms. To cancel the linear term in the presence of the
exponential one is difficult to do numerically. That is why for this
particular system it is convenient to start shooting from large
$\tau$, since both singular solutions at zero share the same
behavior $(\sim 1/\tau)$.

\end{document}